\documentclass[12pt,letterpaper]{article}
\usepackage{amsmath}
\usepackage{mathtools} 
\usepackage{amssymb}
\usepackage{amsthm}
\usepackage{bbm}
\usepackage{graphicx,epsf}
\usepackage{epstopdf}
\usepackage{enumerate}
\usepackage{caption}
\usepackage{natbib}
\usepackage{bm}

\DeclareMathOperator{\ave}{\mbox{ave}}
\DeclareMathOperator{\std}{\mbox{stdev}}
\DeclareMathOperator{\median}{\mbox{median}}
\DeclareMathOperator{\MAD}{\mbox{MAD}}
\DeclareMathOperator{\sign}{\mbox{sign}}
\DeclareMathOperator{\PM}{\mbox{PM}}
\DeclareMathOperator{\Var}{\mbox{Var}}
\DeclareMathOperator{\Std}{\mbox{Stdev}}
\DeclareMathOperator{\Cov}{\mbox{Cov}}
\DeclareMathOperator{\Cor}{\mbox{Cor}}
\DeclareMathOperator{\bias}{\mbox{bias}}
\DeclareMathOperator{\MSE}{\mbox{MSE}}

\newcommand{\eps}{\varepsilon}
\newcommand{\gs}{\geqslant}
\newcommand{\ls}{\leqslant}
\newcommand{\tx}{x^*}
\newcommand{\ty}{y^*}

\newcommand{\bmu}{\boldsymbol \mu}
\newcommand{\br}{\boldsymbol r}
\newcommand{\bt}{\boldsymbol t}

\newcommand{\bv}{\boldsymbol v}
\newcommand{\bx}{\boldsymbol x}

\newcommand{\bC}{\boldsymbol C}
\newcommand{\bX}{\boldsymbol X}
\newcommand{\bY}{\boldsymbol Y}
\newcommand{\bZ}{\boldsymbol Z}
\newcommand{\hmu}{\hat{\mu}}
\newcommand{\hbmu}{\hat{\boldsymbol \mu}}
\newcommand{\hs}{\hat{\sigma}}
\newcommand{\hbs}{\hat{\boldsymbol \Sigma}}

\newcommand{\VF}{V_F}

\newtheorem{proposition}{Proposition}

\newtheorem{corollary}{Corollary}
\newtheorem{definition}{Definition}

\begin{document}

\def\spacingset#1{\renewcommand{\baselinestretch}%
{#1}\small\normalsize} \spacingset{1}


\title{\bf Fast robust correlation\\ 
	           for high-dimensional data}		
\author{Jakob Raymaekers and Peter J. 
	  Rousseeuw\thanks{
	  This research	has been supported by 
		projects of Internal Funds KU Leuven.} 
		\vspace{.4cm} \\
		Department of Mathematics, KU Leuven, Belgium}	
\date{October 20, 2019}
\maketitle

\bigskip
\begin{abstract}
The product moment covariance matrix is 
a cornerstone
of multivariate data analysis, from which one
can derive correlations, principal components,
Mahalanobis distances and many other results.
Unfortunately the product moment covariance and
the corresponding Pearson correlation are very 
susceptible to outliers (anomalies) in the data. 
Several robust estimators of covariance 
matrices have been developed, 
but few are suitable for the ultrahigh dimensional 
data that are becoming more prevalent nowadays.
For that one needs methods whose computation scales well 
with the dimension, are guaranteed to yield a positive
semidefinite matrix, and are sufficiently 
robust to outliers as well as sufficiently accurate in 
the statistical sense of low variability.
We construct such methods using data transformations.
The resulting approach is simple, fast and widely
applicable.
We study its robustness by deriving influence 
functions and breakdown values, and computing the
mean squared error on contaminated data.
Using these results we select a method that performs 
well overall. 
This also allows us to construct a faster 
version of the DetectDeviatingCells method
\citep{Rousseeuw:DDC}
to detect cellwise outliers, that can deal with much 
higher dimensions.
The approach is illustrated on genomic data with
12,600 variables and color video data with
920,000 dimensions.
\end{abstract}

\thispagestyle{empty}

\vskip0.3cm
\noindent
{\it Keywords:} anomaly detection, cellwise outliers,
covariance matrix, data transformation, distance 
correlation.\\

\spacingset{1.45} 
\section{Introduction}
\label{sec:intro}

The most widely used measure of correlation 
is the product-moment correlation coefficient.
Its definition is quite simple. Consider a paired
sample, that is $\{(x_1,y_1),\ldots,(x_n,y_n)\}$\;
where the two numerical variables are the 
column vectors
$X_n = (x_1,\ldots,x_n)^T$ and $Y_n$. Then the
{\it product moment} of $X_n$ and $Y_n$ is
just the inner product
\begin{equation}\label{eq:PM}
  \PM(X_n,Y_n) \;=\;
	  \frac{1}{n} \big\langle X_n,Y_n 
		  \big\rangle \;=\; 
		\frac{1}{n} X_n^T Y_n \;=\; 
		\ave_{i=1}^n x_i y_i \;\; .
\end{equation}
When the $(x_i,y_i)$ are i.i.d. observations of a
stochastic vector $(X,Y)$ the population version 
is the expectation $E[XY]$.
The product moment \eqref{eq:PM} lies at the basis 
of many concepts.
The {\it empirical covariance} of $X_n$ and 
$Y_n$ is the `centered' product moment
\begin{equation}\label{eq:Cov}
  \Cov(X_n,Y_n) \;=\; \frac{n}{n-1} 
	\PM(X_n - \ave(X_n),Y_n - \ave(Y_n))
\end{equation}
with population version $E[(X-E[X])(Y-E[Y])]$\;.
Therefore \eqref{eq:PM} can be seen as a 
`covariance about zero'.
And finally, the product-moment correlation 
is given by
\begin{equation}\label{eq:Pearson}
  \Cor(X_n,Y_n) \;=\;  
	 \frac{n}{n-1} \PM(z(X_n),z(Y_n))
\end{equation}
where the z-scores are defined as
$z(X_n) = (X_n - \ave(X_n))/\Std(X_n)$
with the standard deviation 
   $\Std(X_n) = \sqrt{\Var(X_n)}
    = \sqrt{\Cov(X_n,X_n)}$\;.

The product-moment quantities
\eqref{eq:PM}--\eqref{eq:Pearson}
satisfy
$\PM(X_n,Y_n) = \PM(Y_n,X_n)$ and\newline
$\PM(X_n,X_n) \gs 0$\;.
They have several nice properties.
The {\bf independence property} states that
when $X$ and $Y$ are independent we have
$\Cov(X,Y) = 0$ (assuming the variances exist).
Secondly, when our data set $\bX_{n,d}$ has 
$n$ rows (cases) and $d$ columns (variables, 
dimensions) we can assemble all the product 
moments between the 
variables in a $d \times d$ matrix
\begin{equation}\label{eq:PMmatrix}
  \PM(\bX_{n,d}) = 
  \frac{1}{n} \bX_{n,d}^T \bX_{n,d} \;\; .
\end{equation}
The {\bf PSD property} says
that the matrix \eqref{eq:PMmatrix} is 
positive semidefinite, which is crucial. 
For instance, we can carry out 
a spectral decomposition of the covariance (or 
correlation) matrix, which forms the basis of
principal component analysis. When $d < n$
the covariance matrix will typically be positive
definite hence invertible, which is essential for
many multivariate methods such as the Mahalanobis 
distance and discriminant analysis.
The third property is {\bf speed}: 
the product moment, covariance and correlation 
matrices can be computed very fast, even in high 
dimensions $d$.

Despite these attractive properties, it has been known
for a long time that the product-moment covariance 
and correlation are overly sensitive to outliers
in the data.
For instance, adding a single far outlier can 
change the correlation from $0.9$ to zero or to 
$-0.9$. 

Many robust alternatives to the Pearson correlation 
have been proposed in order to reduce the effect of 
outliers. The first one was probably Spearman's
(1904) correlation coefficient, in which the $x_i$
and $y_i$ are replaced by their ranks. 
Rank-based correlations do not measure a
linear relation but rather a monotone one, which may 
or may not be preferable in a given application.

A second approach is based on the identity
\begin{equation}\label{eq:GK}
  \Cor(X,Y) = 
  \frac{\Var(\tilde{X}+\tilde{Y})-
	      \Var(\tilde{X}-\tilde{Y})}
       {\Var(\tilde{X}+\tilde{Y})+
			  \Var(\tilde{X}-\tilde{Y})}
\end{equation}
where $\tilde{X}=X/\sqrt{Var(X)}$ and
$\tilde{Y}=Y/\sqrt{Var(Y)}$.
\cite{Gnanadesikan:RobEst} proposed to 
replace the nonrobust variance by a robust scale
estimator. This approach is quite popular, see 
e.g. \citep{Shev:book}.  
It does not satisfy the independence property 
however, and the resulting correlation 
matrix is not PSD so it needs to be 
orthogonalized, 
yielding the OGK method of \cite{Maronna:OGK}.

Thirdly, one can start by computing a robust 
covariance matrix $\bC$ such as 
the Minimum Covariance Determinant (MCD) 
method of 
\cite{Rousseeuw:LMS}.
Then we can define a robust correlation
measure between variables $X_j$ and $X_k$ 
by
\begin{equation}\label{eq:Cov2Cor}
  R(X_j,X_k) := 
  C_{jk}/\sqrt{C_{jj}C_{kk}}\;\;.
\end{equation}
In this way we do produce a PSD 
matrix, but we lose the independence property.
In fact, here the robust correlation between 
two variables depends on the other variables, 
so adding or removing a variable changes it.
Also, the computational requirements 
do not scale well with the dimension $d$, 
making this approach infeasible for 
high dimensions.

Another possibility is to start from the
Spatial Sign Covariance Matrix (SSCM) of
\cite{Visuri:Rank}.
This method first computes the 
{\it spatial median} $\hbmu$ of the data 
points $\bx_i$ 
by minimizing $\sum_i ||\bx_i - \bmu||$. 
It then computes the product moment of the
so-called {\it spatial signs}
$(\bx_i - \hbmu)/||\bx_i-\hbmu||$.
Then \eqref{eq:Cov2Cor} can be applied.
The result is PSD but does not satisfy the
independence property either. 

For high-dimensional data, the 
product-moment technology is computationally
attractive. 
This suggests using the idea
underlying Spearman's rank correlation, which 
is to transform the variables first. 
We do not wish to restrict ourselves to ranks
however, and we want to explore how far 
the principle of robustness by data 
transformation can be pushed.

In general, we consider a transformation $g$ 
applied to the individual variables, and we 
define the resulting $g$-product moment as
\begin{equation}\label{eq:PMg}
   \PM_g (X_n,Y_n) \;\;:=\;\;
   \PM(g(X_n),g(Y_n))
\end{equation}
and similarly for $\Cov_g$ and $\Cor_g$.
Choosing $g(x_i)=x_i$ yields the usual product 
moment, and setting $g(x_i)$ equal to its rank 
yields the Spearman correlation. 
The $g$-product moment approach satisfies 
all three desired properties.
First of all, if we use a bounded function 
$g$ the population version $E[g(X)g(Y)]$ always 
exists and $\Cov_g$ satisfies the independence 
property without any moment conditions.
Secondly, the resulting matrices 
$\PM_g(\bX_{n,d}) = 
 \PM(g(X_{.1}),\ldots,g(X_{.d}))$ 
always satisfy the PSD property. 
And finally, this method is very fast
provided the transformation $g$ can be 
computed quickly (which could even be done in 
parallel over variables).

Note that the bivariate winsorization in 
\cite{Khan:RLARS} is a transformation
$\tilde{g}(X_n,Y_n)$ that depends on both
arguments simultaneously, unlike \eqref{eq:PMg}.
It yields a good robust bivariate correlation 
but without the multivariate PSD property.

Our present goal is to find transformations $g$
for \eqref{eq:PMg} that yield covariance matrices 
that are 
sufficiently robust and at the same time 
sufficiently efficient in the statistical sense.

\begin{table}[ht]
\small
\centering
\caption{Computation times (in seconds) of 
various correlation matrices as a 
function of the dimension $d$, for $n=1000$
observations.}
\label{tab:times}
\begin{tabular}{|c|cccccc|}
\hline
 dimension & \;\;\;MCD\;\; & \;\;OGK\;\; & \;SSCM\; & 
   Spearman & Wrapping & Classic \\ 
\hline
   10 & 0.319 & 0.022 & 0.004 & 0.002 & 0.003 & 0.001 \\
   50 & 6.222 & 0.426 & 0.009 & 0.009 & 0.012 & 0.002 \\
  100 & 24.76 & 2.089 & 0.031 & 0.019 & 0.027 & 0.008 \\
  500 & 1599  & 44.78 & 0.678 & 0.226 & 0.281 & 0.171 \\
 1000 & -     & 166.7 & 3.107 & 0.774 & 0.836 & 0.685 \\
 5000 & -     & 4389  & 129.1 & 17.11 & 17.39 & 16.81 \\
10000 & -     & -     & 568.9 & 68.24 & 68.78 & 67.27 \\
20000 & -     & -     & 2448  & 278.4 & 274.9 & 273.6 \\
\hline
\end{tabular}
\vskip0.3cm
\end{table} 

Table \ref{tab:times} lists some computation
times (in seconds) of the robust correlation 
methods mentioned above for $n=1000$ generated 
data points in various dimensions $d$, as well
as the classical correlation matrix. 
(The times were measured
on a laptop with Intel Core i7-5600U CPU at
2.60 GHz.) The fifth column is the $g$-product 
moment method that will be proposed in this paper.
Note that the MCD cannot be computed when
$d \geq n$, and that the computation times
of MCD and OGK become infeasible at high 
dimensions.
The next three methods are faster, and their
robustness will be compared later on.

The remainder of the paper is organized as follows. 
In Section \ref{sec:meth} we explore the properties
of the $g$-product moment approach by means of 
influence functions,
breakdown values and other robustness tools, 
and in Section \ref{sec:wrap} we design a new 
transformation $g$ based on what we have learned.
Section \ref{sec:sim} compares these transformations 
in a simulation study and makes recommendations.
Section \ref{sec:highdim} explains how to use
the method in higher dimensions, illustrated
on some real high-dimensional data sets in
Section \ref{sec:app}.

\section{General properties of $g$-product moments}
\label{sec:meth}

The oldest type of robust $g$-product moments 
occur in rank correlations. 
Define a rescaled version of the sample ranks 
as $R_n(x_i) = (\mbox{Rank}(x_i)-0.5)/n$ 
where $\mbox{Rank}(x_i)$ denotes the rank of 
$x_i$ in $\{x_1,\ldots,x_n\}$. 
The population version of $R_n(x_i)$ is the 
cumulative distribution function (cdf) of $X$.
Then the following functions $g$ define 
rank correlations:
\begin{itemize}
\item $g(x_i) = R_n(x_i)$ yields the Spearman rank 
      correlation \citep{Spearman:cor}.
\item $g(x) = \sign(R_n(x_i) - 0.5)$ gives the
      quadrant correlation.
\item $g(x) = \Phi^{-1}(R_n(x))$ (where $\Phi$ is 
      the standard Gaussian cdf) yields the normal 
			scores correlation.
\item $g(x) := \Phi^{-1}\left([R_n(x)]_{\alpha}^
      {1-\alpha}\right)$ with the notation
      $[y]_{a}^{b} := 
			\mbox{min}(b,\mbox{max}(a,y))$
      is the truncated normal
      scores function, first proposed on pages 
			210--211 of \citep{Hampel:IFapproach} 
      in the context of univariate rank tests.
\end{itemize}
Kendall's tau is of a somewhat
different type as it replaces each variable $X_n$
by a variable with $n(n-1)/2$ values, but we
compare with it in Section \ref{sec:sim}.

A second type of robust $g$-product moments 
goes back to Section 8.3 in the book 
of \cite{Huber:RobStat}
and is based on M-estimation.
Huber transformed $x_i$ to
\begin{equation}\label{eq:psitrans}
   g(x_i) = \psi((x_i - \hmu)/\hs)\;,
\end{equation} 
where $\hmu$ is an M-estimator of 
location defined by
$\sum_i \psi((x_i - \hmu)/\hs) = 0$
and
$\hs$ is a robust scale estimator such as
the MAD given by 
$\MAD(X_n) = 
 1.4826\,\median_i|x_i - \median_j(x_j)|$\;.
Note that $(x_i - \hmu)/\hs$ is like a z-score
but based on robust analogs of the mean and 
standard deviation.
For $\psi(z)=\sign(z)$ this yields 
$\hmu = \median_j(x_j)$ so we recover the 
quadrant correlation.
Another transformation is Huber's
$\psi_b$ function given by
$\psi_b(z) = [z]_{-b}^{b}$
for a given corner point $b>0$. 
One can also use the sigmoid transformation
$\psi(z) = \tanh\left(z\right)$.
Note that the transformation \eqref{eq:psitrans}
does not require any tie-breaking rules,
unlike the rank correlations.
\cite{Huber:RobStat} derived the asymptotic
efficiency of the $\psi$-product moment. 
We go further by also computing
the influence function, the breakdown value and
other robustness measures.
Our goal is to find a function $\psi$ that is
well-suited for correlation.

\subsection{Influence function and efficiency}
\label{sec:IF}

Note that the $g$-product moment 
$\PM_g(X_j,X_k)$ between two variables $X_j$ 
and $X_k$ in a multivariate data set does not
depend on the other variables, so we can study 
its properties in the bivariate setting.

For analyzing the statistical properties of the
$\psi$-product moment we assume a simple model for 
the `clean' data, before outliers are added.
The model says that $(X,Y)$ follows a bivariate
Gaussian distribution $F_\rho$ given by
\begin{equation}\label{eq:Frho}
  F_{\rho} = N \left(
\begin{bmatrix}
0 \\
0
\end{bmatrix}, 
\begin{bmatrix}
1 & \rho \\
\rho & 1
\end{bmatrix}
  \right) 
\end{equation}
for $-1 < \rho < 1$, so $F_0$ is just the bivariate
standard Gaussian distribution.
We restrict ourselves to odd functions $\psi$
so that $E[\psi(X)]=0=E[\psi(Y)]$, and study the
statistical properties of
$T_n = \frac{1}{n} \sum_{i=1}^n 
      \psi(x_i)\psi(y_i)$
with population version
   $T_{\psi} = E[\psi(X)\psi(Y)]$.
Note that $T_{\psi}$ maps the bivariate 
distribution of $(X,Y)$ to a real number,
and is therefore called a {\it functional}.
It can be seen as the limiting case of the
estimator $T_n$ for $n \rightarrow \infty$.
On the other hand, a finite sample
$Z_n = \{(x_1,y_1),\ldots,(x_n,y_n)\}$ yields 
an empirical distribution 
 $F_n(x,y) = \frac{1}{n} \sum_{i=1}^n 
      I(x_i \le x,\,y_i \le y)$
and we can define an estimator $T_n(Z_n)$ as
$T_\psi(F_n)$, so there is a strong
connection between estimators and functionals.
Whereas the usual 
consistency of an estimator $T_n$ requires that
$T_n$ converges to $\rho$ in probability,
there exists an analogous notion for 
functionals: $T_\psi$ is called 
{\it Fisher-consistent} for $\rho$ iff 
$T_\psi(F_\rho) = \rho$.
	
We will start with the influence function (IF) of 
$T_{\psi}$. Following \cite{Hampel:IFapproach}, 
the raw influence function of the functional
$T_{\psi}$ at $F_\rho$ is 
defined in any point $(x,y)$ as 
\begin{equation} \label{eq:rawIF}
   \mbox{IF}_{raw}((x,y),T_{\psi},F_\rho) = 
	\frac{\partial}{\partial \eps}
	T_{\psi}((1-\eps)F_\rho + \eps
	\Delta_{(x,y)})|_{\eps = 0}
\end{equation}
where $\Delta_{(x,y)}$ is the probability
distribution that puts all its mass in $(x,y)$.
Note that \eqref{eq:rawIF} is well-defined 
because $(1-\eps)F_\rho + \eps\Delta_{(x,y)}$
is a probability distribution so $T_\psi$
can be applied to it.
The IF quantifies the effect of a small amount 
of contamination in $(x,y)$ on $T_{\psi}$
and thus describes the effect of an outlier 
on the finite-sample estimator $T_n$. 
It is easily verified that 
$\mbox{IF}_{raw}((x,y),T_{\psi},F_0) = 
  \psi(x)\psi(y)$.

However, we cannot compare the raw influence 
function \eqref{eq:rawIF} across different 
functions $\psi$ since $T_{\psi}$ is not 
Fisher-consistent, that is, 
$T_{\psi}(F_\rho) \neq \rho$ in general.
For non-Fisher-consistent statistics $T$ we 
follow the approach of 
\cite{Rousseeuw:IFgeneral} and 
\cite{Hampel:IFapproach} by defining 
\begin{equation}\label{eq:xi}
  \xi(\rho) := T(F_\rho) 
  \;\;\; \mbox{ and } \;\;\;
  U(F) := \xi^{-1}(T(F))
\end{equation}	
so $U$ is Fisher-consistent, and putting
\begin{equation}\label{eq:generalIF}
  \mbox{IF}((x,y),T,F) := 
  \mbox{IF}_{raw}((x,y),U,F) = 
	\frac{\mbox{IF}_{raw}((x,y),T,F)}
	{\xi'(\rho)}\;\;.
\end{equation}

\begin{proposition}\label{prop:IF}
When $\psi$ is odd [i.e. 
$\psi(-z)=-\psi(z)$] and bounded we have  
$\xi'(0) = E[\psi']^2$ hence the 
influence function of $T_{\psi}$ at 
$F_0$ becomes
\begin{equation}\label{eq:IFT}
  \mbox{IF}((x,y),T_{\psi},F_0) =
  \frac{\psi(x)\psi(y)}{E[\psi']^2}.
\end{equation}
\end{proposition}
\noindent The proof can be found in 
Section \ref{A:proofIFT} of the 
Supplementary Material. The influence function
at $F_\rho$ for $\rho \neq 0$ derived in 
Section \ref{A:IFgen} has the same overall 
shape.

Since the IF measures the effect of outliers
we prefer bounded $\psi$, unlike the classical 
choice $\psi(z) = z$. 
Note that \eqref{eq:IFT} is the raw
influence function of 
 $T^{*} = E[\psi^{*}(X)\psi^{*}(Y)]$ 
at $F_0$, where $\psi^{*}(u) = \psi(u)/E[\psi']$.
As $\psi$ is bounded $T^{*}$ is integrable,
so by the law of large numbers $T_n^{*}$ is 
strongly consistent for its functional value:
$T_n^* =
  \frac{1}{n}\sum_{i=1}^{n}
	   {\psi^*(x_i)\psi^*(y_i)} 
  \xrightarrow{a.s.} 
	T^{*}(F_{\rho})$  for $n \to \infty$.
By the central limit theorem, $T^{*}$ is then
asymptotically normal under $F_0$:
\begin{equation*}
\sqrt{n}(T_n^{*}-0)\rightarrow N(0,V)\;,
\end{equation*}
where 
\begin{equation}\label{eq:V}
  V = \frac{E[\psi^2]^2}{E[\psi']^4} =
	\left(\frac{E[\psi^2]}{E[\psi']^2}\right)^2.
\end{equation}
From this we obtain the asymptotic efficiency 
$\mbox{eff} = (E[\psi']^2/E[\psi^2])^2$\;.

Note that the influence function of $T_{\psi}$ at 
$F_0$ factorizes as the product of the influence 
functions of the M-estimator $L_{\psi}$ of 
location with the same $\psi$-function:
\begin{equation}\label{eq:splitinf}
  \mbox{IF}((x,y),T_{\psi},F_0) = 
 	\mbox{IF}(x,L_{\psi},\Phi)\,
	\mbox{IF}(y,L_{\psi},\Phi)\;,
\end{equation}
because $\mbox{IF}(x,L_{\psi},\Phi) =
\psi(x)/E[\psi']$\,.
This explains why the efficiency of $T_{\psi}$ 
satisfies $\mbox{eff}(T_{\psi}) = 
 (\mbox{eff}(L_{\psi}))^2$\;. 
We are also interested in attaining a low
gross-error sensitivity $\gamma^*(T_{\psi})$, 
which is defined as the supremum 
of $|\mbox{IF}((x,y),T_{\psi},F_0)|$ and 
therefore equals $(\gamma^*(L_{\psi}))^2$\;.
It follows from \citep{Rousseeuw:CVC} that the 
quadrant correlation $\psi(z) = \sign(z)$ has 
the lowest gross-error sensitivity among all 
statistics of the type 
  $T_{\psi} = E[\psi(X)\psi(Y)]$. 
In fact,
$\mbox{IF}((x,y),T_{\psi},F_0) = 
(\pi/2) \sign(x)\sign(y)$
yielding $\gamma_{T}^{*} = \pi/2$.
However, the quadrant correlation is very 
inefficient as $\mbox{eff} = 4/\pi^2 = 40.5\%$.

The influence functions of rank correlations
are obtained by \cite{Croux:IFspearman} and
\cite{Boudt:GRcor}. 
Note that for some rank correlations the 
function $\xi$ of \eqref{eq:xi} is known 
explicitly, in fact
$\xi(\rho) = \sin(\rho \pi/2)$ 
for the quadrant correlation, 
$\xi(\rho) = (6/\pi)\arcsin(\rho/2)$ 
for Spearman and 
$\xi(\rho)= \rho$ 
for normal scores.
It turns out that these IF at $F_0$ match the 
expression in Proposition \ref{prop:IF} if
$\psi$ corresponds to the population version
of the transformation $g$ in the rank 
correlation, as explained
in Section \ref{A:rankIF} of the 
Supplementary Material.

\begin{figure}[!ht]
\centering
\includegraphics[width=0.7\textwidth]
  {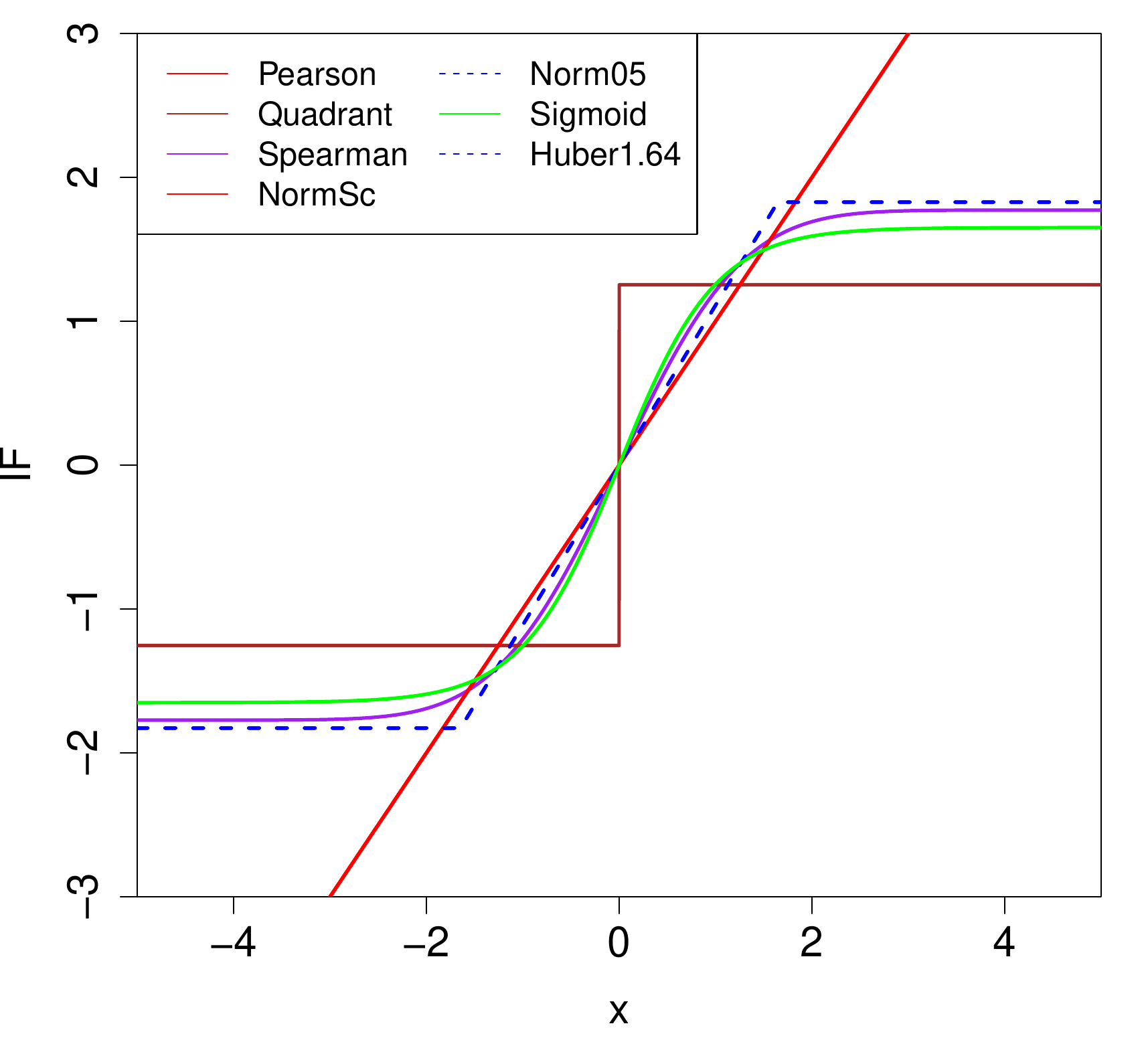}
\vskip-0.2cm					
\caption{Location influence functions at 
  $\rho=0$ for different transformations $g$}
\label{fig:IF}
\end{figure}

The influence functions of rank correlations 
at $F_0$ also factorize as in 
\eqref{eq:splitinf}.
Figure \ref{fig:IF} plots these location
influence functions for several choices 
of the transformation $g$. 
We see that the Pearson and normal scores
correlations have the same influence function
(the identity), which is unbounded.
On the other hand, the IF of Huber's $\psi_b$
stays constant outside the corner points $-b$
and $b$.
The truncated normal scores (`Norm05') has 
the same IF as Huber's $\psi_b$ provided 
$\alpha = \Phi(-b)$\;.
The Spearman rank correlation and the sigmoid
transformation have smooth influence functions.

\subsection{Maxbias and breakdown value}

Whereas the IF measures the effect of one or
a few outliers, we are now interested in the
effect of a larger fraction $\eps$ of 
contamination.
For the uncontaminated distribution of the 
bivariate $(X,Y)$ we take the Gaussian 
distribution $F=F_\rho$ given
by \eqref{eq:Frho}. 
Then we consider all contaminated distributions
of the form
\begin{equation}\label{eq:epscont}
  F_{H,\eps} = (1-\eps)F+\eps H\;,
\end{equation}
where $\eps \gs 0$ and $H$ can be any 
distribution.
This {\it $\eps$-contamination model} is 
similar to the contaminated distributions in 
\eqref{eq:rawIF} and  \eqref{eq:CVC}
but here $H$ is more general.

A fraction $\eps$ of contamination can induce
a maximum possible upward and downward bias on 
$T_\psi = \Cor(\psi(X),\psi(Y))$
denoted by
\begin{equation}\label{eq:maxbias}
   B^+(\eps,T_\psi,F) =
	 \sup_{G \in \mathcal{F}_\eps} 
	 (T_\psi(G)-T_\psi(F))
	 \;\; \mbox{ and } \;\;
   B^-(\eps,T_\psi,F) =
	 \inf_{G \in \mathcal{F}_\eps}
	 (T_\psi(G)-T_\psi(F))\;,
\end{equation}
where $\mathcal{F}_\eps =
 \{G;\; G = (1-\eps)F +\eps H\;\; 
  \mbox{for any distribution }H\}$\;.
The proof of the following proposition is 
given in Section \ref{A:proofbias} in the 
Supplementary Material.
\begin{proposition}\label{prop:corbias}
Let $\eps\in [0,1]$ be fixed and $\psi$ be
odd and bounded. 
Then the maximum upward bias of $T_\psi$
at $F$ is given by 
\begin{equation}
  B^+(\eps,T_\psi,F) = \frac{(1-\eps)
	\Var_F(\psi(X))\,T_\psi(F) + \eps M^2}
	{(1-\eps)\Var_F(\psi(X)) + \eps M^2}
	-T_\psi(F)
\end{equation}
with $M := \sup_x |\psi(x)|$, and
the maximum downward bias is 
\begin{equation}
B^-(\eps,T_\psi,F) = \frac{(1-\eps)
	\Var_F(\psi(X))\,T_\psi(F) - \eps M^2}
	{(1-\eps)\Var_F(\psi(X)) + \eps M^2}
	-T_\psi(F)\;\;.
\end{equation}
\end{proposition}

\vskip0.1cm
The {\it breakdown value} $\eps^*$ of a 
robust estimator is loosely defined as the 
smallest $\eps$ that can make the result 
useless.
For instance, a location estimator $\hmu$
becomes useless when its maximal bias tends to
infinity.
But correlation estimates stay in the bounded 
range $[-1,1]$ hence the bias can never exceed 
2 in absolute value, so the situation is not as
clear-cut and several alternative definitions
could be envisaged.
Here we will follow the approach of
\cite{Caperaa:tauxderes} who
define the breakdown value of a correlation 
estimator as the smallest amount of 
contamination needed to give perfectly 
correlated variables a negative correlation. 
More precisely:
\begin{definition}\label{def:bdp}
Let $F$ be a bivariate distribution with 
$X=Y$, and $R$ be a correlation measure. 
Then the breakdown value of $R$ 
is defined as
\begin{equation*}
\eps^{*}(R) = \inf \{\eps > 0\; ; \; 
   \inf_{G \in \mathcal{F}_\eps}
	 R(G) \ls 0\}\;\;.							
\end{equation*}
\end{definition}

The breakdown value of $T_\psi$ then follows 
immediately from 
Proposition \ref{prop:corbias}:

\begin{corollary}\label{prop:breakdown}
When $\psi$ is odd and bounded the breakdown 
value $\eps^{*}$ of $T_\psi$ equals 
\begin{equation*}
  \eps^{*}(T_\psi)=
  \frac{\Var_F(\psi(X))}
	     {\Var_F(\psi(X))+M^2} \;\;.
\end{equation*}
\end{corollary}

The breakdown values of rank correlations 
were obtained in
\citep{Caperaa:tauxderes,Boudt:GRcor}.
They used a different 
contamination model, but their results
still hold under $\eps$-contamination 
as shown in Section \ref{A:rankBD} 
in the Supplementary Material.
  
\section{The proposed transformation}
\label{sec:wrap}

The change-of-variance curve
\citep{Hampel:tanh,Rousseeuw:CVC} is 
given by
\begin{equation}\label{eq:CVC}
\mbox{CVC}(z,T_{\psi},F) = \frac{\partial}
  {\partial \eps}
  \left[ \log V\big(T_{\psi}, (1-\eps)F +
	\eps(\Delta_z + \Delta_{-z})/2\big) 
	\right]|_{\eps = 0}
\end{equation}
and measures how stable the variance 
of the method is when the underlying 
distribution is contaminated, which may
make it longer tailed.
We do not want the variance to grow too much,
as is measured by the change-of-variance 
sensitivity $\kappa^{*}(T_{\psi})$, 
which is the supremum of the CVC.
(On the other hand, negative values of
the CVC indicate lower variance and are
not a concern.)
Since the asymptotic variance of $T_{\psi}$
satisfies $V(T_{\psi}) = (V(L_{\psi}))^2$ 
we obtain 
$\mbox{CVC}(z,T_{\psi},F_0) =
 2\,\mbox{CVC}(z,L_{\psi},\Phi)$ and
$\kappa^*(T_{\psi}) = 
 2\,\kappa^*(L_{\psi})$\;.
Therefore we inherit all the results 
about the CVC from the location setting. 
For instance, the quadrant correlation 
[with $\psi(z) = \sign(z)$] has 
the lowest possible $\kappa^*(T_{\psi})$\;. 

Now suppose one wants to eliminate the
effect of far outliers, say those that lie 
more than $c$ robust standard deviations away. 
This can be done by imposing 
\begin{equation}\label{eq:redesc}
  \psi(z)=0 \;\;\; \mbox{whenever}
	          \;\;\; |z| > c \;\;.
\end{equation}
Such functions $\psi$ can no longer be 
monotone, and are called {\it redescending} 
instead.
They were first used for M-estimation of 
location, and performed extremely well in 
the seminal simulation study of 
\cite{Andrews:1972}. They have been used
in M-estimation ever since.

\begin{figure}[!ht]
\centering
\includegraphics[width=0.6\textwidth]
                {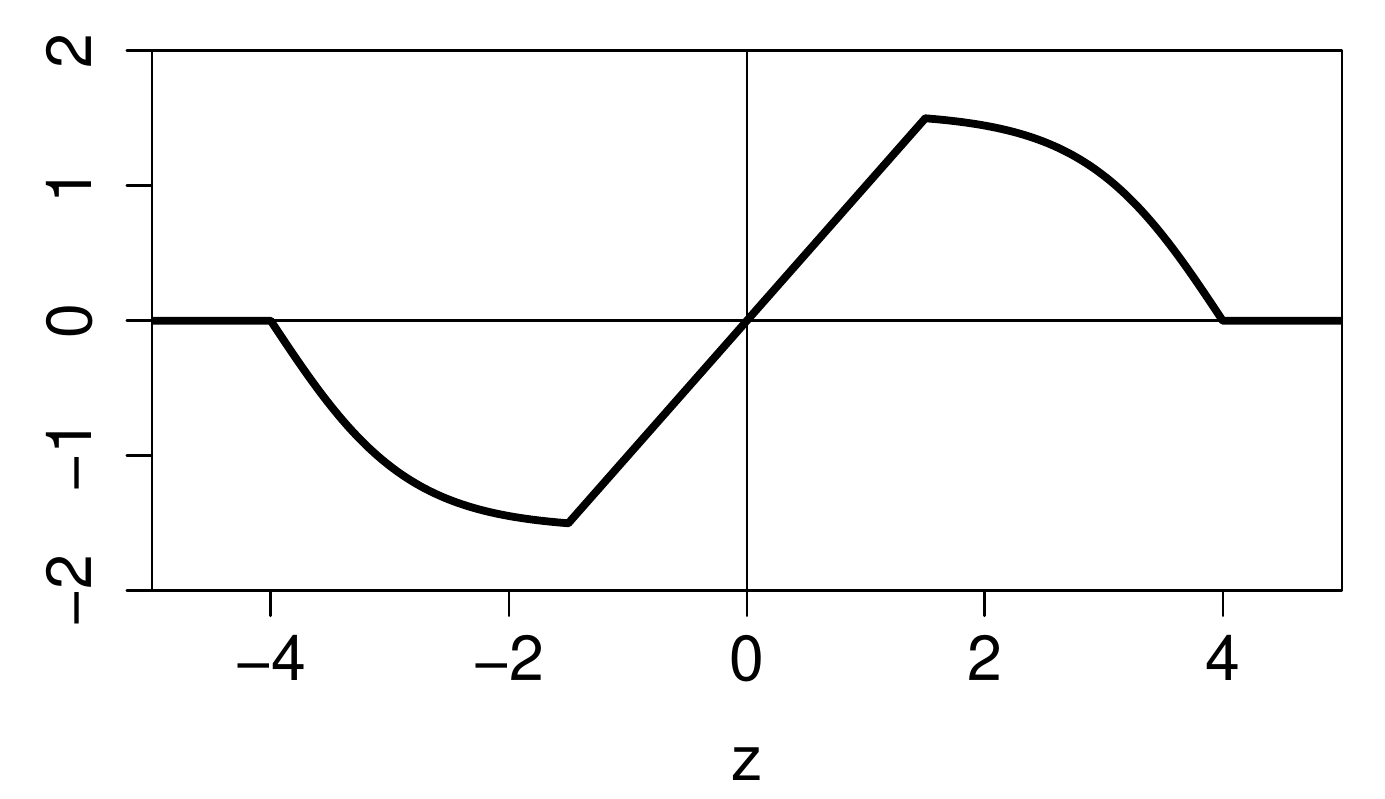}
\vskip-0.2cm					
\caption{The proposed transformation 
 	\eqref{eq:psiwrap}
	with default constants $b=1.5$ and $c=4$.}
\label{fig:psiwrap}
\end{figure}

In the context of location estimation,
\cite{Hampel:tanh} show that the  
$\psi$-function satisfying \eqref{eq:redesc} 
with the highest efficiency subject to a 
given $\kappa^*(T_{\psi})$ is of the 
following form:
\begin{equation}\label{eq:psiwrap}
  \psi_{b,c}(z) = \begin{cases}
  z & \mbox{ if } 0 \ls |z| \ls b\\
  q_1 \tanh\big(q_2(c-|z|)\big) \sign(z)
	& \mbox{ if } b \ls |z| \ls c \\
	0 & \mbox{ if } c \ls |z|\;\;.
\end{cases}
\end{equation}
For any combination $0<b<c$ the values of 
$q_1$ and $q_2$ can be derived as in Section 
\ref{A:wrapping} of the Supplementary Material.
Our default choice is $b=1.5$ and $c=4$ 
as in Figure \ref{fig:psiwrap}.
As we will see in Table \ref{tab:corrs}
this choice strikes a good compromise between 
robustness and efficiency.
Note that the $b$ in $\psi_{b,c}$ plays the same 
role as the ``corner value'' in the Huber 
$\psi_b$ function for location estimation. 
In that setting, $b = 1.5$ has been a popular choice 
from the beginning. The value $c=4$ reflects that we 
do not trust measurements that lie more than 4 
standard deviations away.
The form of $\psi_{b,c}(z)$ for $b \ls |z| \ls c$ 
is the result of solving a differential equation.

A nice property of $\psi_{b,c}$ is that 
under normality a large 
majority of the data values
(in fact $86.6\%$ of them for $b=1.5$)
are left unchanged by the transformation, 
and only a minority is modified. 
Leaving the majority of the data unchanged has
the advantage that we keep much information
about the distribution of a variable and the
type of association between variables (e.g.
linear), unlike rank transforms.

\begin{figure}[!ht]
\centering
\includegraphics[width=0.7\textwidth]
      {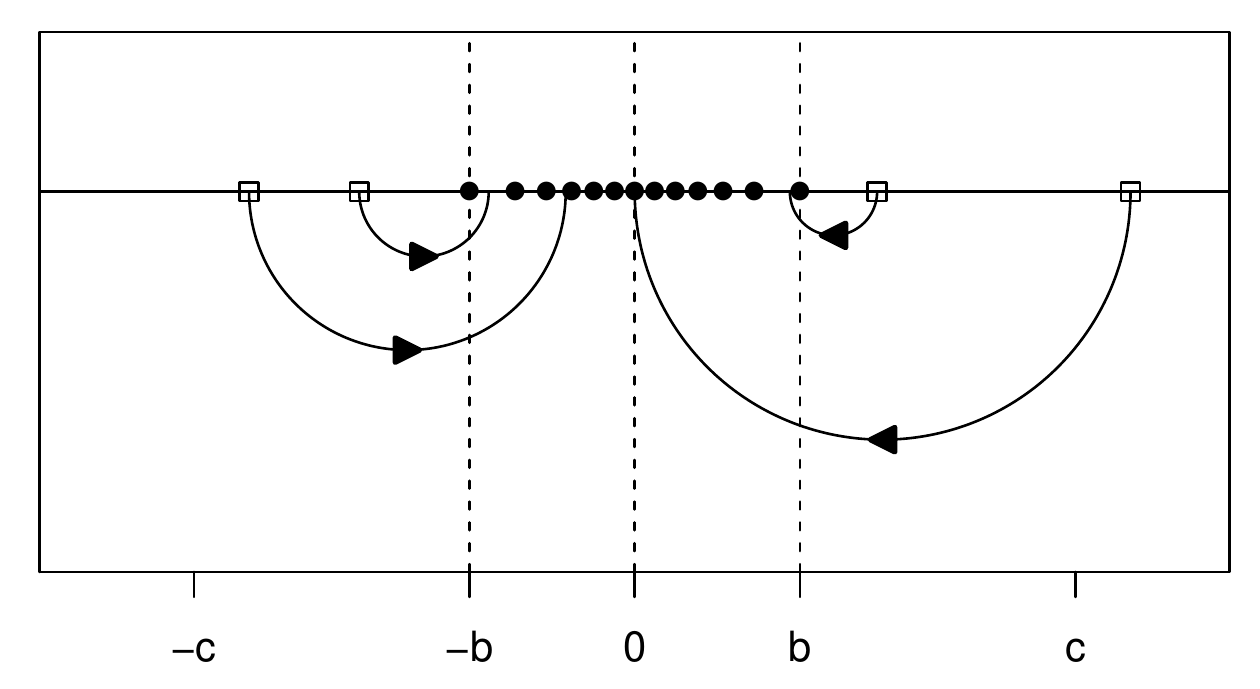}
\vskip-0.2cm					
\caption{Illustration of wrapping
  a standardized sample
	$\{z_1,\ldots,z_n\}$\;.
  Values in the interval $[-b,b]$ are left
	unchanged, whereas values outside
	$[-c,c]$ are zeroed. The intermediate
	values are `folded' inward so they still
	play a role.}
\label{fig:wrapping}
\end{figure}

Interestingly, $\psi_{b,c}$ pushes values 
between $b$ and $c$ closer to the center so
intermediate outliers still play some smaller 
role in the correlation, whereas far outliers 
do not count.
For this reason we refer to $\psi_{b,c}$ 
as the {\it wrapping function},
as it wraps the data around the interval
$[-b,b]$\,. Indeed, the points on the interval
are mapped to themselves, whereas the other
points are wrapped around the
corners, as in Figure \ref{fig:wrapping}.

Another way to describe this is to
say that wrapping multiplies the variable
$z$ by a weight $w(z)$, where
$w(z) \coloneqq 1$ when $|z| \le b$ and 
$w(z) \coloneqq \psi_{b,c}(z)/z$ for 
$|z| > b$.

The influence function \eqref{eq:splitinf}
contains $\mbox{IF}(z,L_{\psi},\Phi) =
\psi_{b,c}(z)/E[\psi'_{b,c}]$, which has
the shape of $\psi_{b,c}$ in Figure
\ref{fig:psiwrap}.
The bivariate influence function
$\mbox{IF}((x,y),T_{\psi},F_\rho)$ 
is continuous and bounded, and shown
in Figure \ref{fig:IF05} in
Section \ref{A:wrapping} of the
Supplementary Material.

Table \ref{tab:corrs} lists some 
correlation
measures based on transformations $g$ that
either use ranks or $\psi$-functions. 
For each the breakdown value $\eps^*$ and
the efficiency and gross-error sensitivity 
$\gamma^*$ at $\rho = 0$ are listed.
The rejection point $\delta^*$ says 
how far an outlier must lie before the IF
is zero.
The last column shows
the product-moment correlation between a
Gaussian variable $X$ and its transformed
$g(X)$\,.
The correlation is quite high for most 
transformations studied here, providing
insight as to why this approach works.

\begin{table}[ht]
\centering
\caption{Correlation measures based on
transformations $g$ with their
breakdown value $\eps^*$, 
efficiency, gross-error sensitivity 
$\gamma^*$,
rejection point $\delta^*$ and 
correlation between $X$ and $g(X)$.}
\label{tab:corrs}
\vskip0.3cm
\begin{tabular}{|c|c|c|c|c|c|}
\hline
$\Cor_g$ & $\eps^{*}$ & eff & $\gamma^*$
  & $\delta^*$ & $\mbox{Cor}$\\ 
\hline
\hline
Pearson & 0\% & 100\% & $\infty$ & $\infty$ & 1\\
\hline
Quadrant & 50\% & 40.5\% & 1.57 & $\infty$ & 0.798\\ 
Spearman (SP) & 20.6\% & 91.2\% & 3.14 & $\infty$
  & 0.977\\
Normal scores (NS) & 12.4\% & 100\% & 
  $\infty$ & $\infty$ & 1\\
Truncated NS, $\alpha = 0.05$ & 16.3\% &
  95.0\% & 3.34 & $\infty$ & 0.987\\
Truncated NS, $\alpha = 0.1$ & 20.7\% &
  88.9\% & 2.57 & $\infty$ & 0.971\\
\hline
Sigmoid & 28.3\% & 86.6\% & 2.73 & $\infty$ & 0.965\\
Huber, $b = \Phi^{-1}(0.95) \approx 1.64$
        & 23.5\% & 95.0\% & 3.34 & $\infty$ & 0.987\\
Huber, $b = \Phi^{-1}(0.9) \approx 1.28$
        & 29.2\% & 88.9\% & 2.57 & $\infty$ & 0.971\\
Wrapping, $b=1.5$, $c=4$ & 25.1\% & 89.0\%
   & 3.16 & 4.0 & 0.971\\
Wrapping, $b=1.3$, $c=4$ & 28.1\% & 84.4\%
   & 2.79 & 4.0 & 0.958\\
\hline
\end{tabular}
\vskip0.3cm
\end{table}

In Table \ref{tab:corrs} we see that the 
quadrant correlation has the highest breakdown 
value but the lowest efficiency. 
The Spearman correlation reaches a much better 
compromise between breakdown and efficiency.
Normal scores has the asymptotic efficiency and
IF of Pearson but with a breakdown value of 
12.4\%, a nice improvement.
Truncating 5\% improves its robustness a bit at 
the small cost of 5\% of efficiency, whereas 
truncating 10\% brings its performance close 
to Spearman. 

Both the Huber and the wrapping correlation
have a parameter $b$, the corner point, which
trades off robustness and efficiency.
A lower $b$ yields a higher breakdown value
and a better gross-error sensitivity, but
a lower efficiency.
Note that the Huber correlation looks good in 
Table \ref{tab:corrs}, but in the simulation
study of Section \ref{sec:sim} it performs
less well than wrapping in the presence of 
outliers, and the same holds in the real data
application in Section \ref{sec:video}.
The reason is that wrapping gives a lower
weight $w(z) := \psi_{b,c}(z)/z$ to outliers
and even $w(z) = 0$ for $|z| > c$, whereas the 
Huber weight $w_b(z) := \psi_b(z)/z$ is higher 
for outliers and always nonzero, so even far 
outliers still have an effect.
  
Note that whenever two random variables 
$X$ and $Y$ are
independent the correlation between the wrapped
variables $g_X(X)$ and $g_Y(Y)$ is zero, even 
if the original $X$ and $Y$ did not satisfy any 
moment conditions. This follows from the 
boundedness of $\psi_{b,c}$ in \eqref{eq:psiwrap}. 

It is well-known that the reverse is not true
for the classical Pearson correlation, but that
it holds when $(X,Y)$ follow a bivariate
Gaussian distribution. This is also true for the
wrapped correlation.
\begin{proposition}\label{prop:independence}
If the variables $(X,Y)$ follow a bivariate
Gaussian distribution and the correlation
between the wrapped variables $g_X(X)$ and 
$g_Y(Y)$ is zero, then $X$ and $Y$ are 
independent.
\end{proposition}
Another well-known property says that the
Pearson correlation of a dataset
$Z = \{(x_1,y_1),\ldots,(x_n,y_n)\}$ equals
1 if and only if there are constants $\alpha$ 
and $\beta$ with $\beta>0$ such that
\begin{equation} \label{eq:linear} 
  y_i = \alpha + \beta x_i
\end{equation}
for all $i$ (perfect linear relation).
The wrapped correlation satisfies a similar
result.

\begin{proposition}\label{prop:linearity}
(i) If \eqref{eq:linear} holds for all $i$
    and we transform the data to
$g_X(x_i) = \psi_{b,c}((x_i - \hmu_X)/\hs_X)$
		and 	
$g_Y(y_i) = \psi_{b,c}((y_i - \hmu_Y)/\hs_Y)$ 
    then $\Cor(g_X(x_i),g_Y(y_i)) = 1$.
		
(ii) If  $\Cor(g_X(x_i),g_Y(y_i)) = 1$
     then \eqref{eq:linear} holds for all 
		 $i$ for which 
		 $|x_i - \hmu_X|/\hs_X \leqslant b$ and
		 $|y_i - \hmu_Y|/\hs_Y \leqslant b$.
\end{proposition}

In part (ii) the linearity has to hold for 
all points with coordinates in the central 
region of their distribution, whereas far
outliers may deviate from it.
In that case the points in the central 
region are exactly fit by a straight line.
The proofs of Propositions
\ref{prop:independence} and
\ref{prop:linearity} can be found in
Section \ref{A:independence} of the 
Supplementary Material.

{\bf Remark.} Whereas Proposition
\ref{prop:independence} requires bivariate
gaussianity, the other results in this
paper do not. In fact, Propositions
\ref{prop:IF}, \ref{prop:corbias}, and
\ref{prop:linearity} as well as Corollary
\ref{prop:breakdown} still hold when the
data is generated by a symmetric and unimodal
distribution.
The corresponding proofs in the Supplementary 
Material are for this more general setting.

\section{Simulation Study}
\label{sec:sim}
We now compare the correlation by 
transformation methods in 
Table \ref{tab:corrs} for finite samples.
For all of these methods the correlation
between two variables does not depend on
any other variable in the data, so we only
need to generate bivariate data here.

For the non rank-based methods we first 
normalize each variable by a robust scale
estimate, and then estimate the location
by the M-estimator with the given
function $\psi$.
Next we transform $x_i$ to 
$\tx_i = \psi((x_i - \hmu_X)/\hs_X)$
and $y_i$ to
$\ty_i = \psi((y_i - \hmu_Y)/\hs_Y)$
and compute the plain Pearson correlation
of the transformed sample 
$\{(\tx_1,\ty_1),\ldots,(\tx_n,\ty_n)\}$.

{\bf Clean data.}
Let us start with uncontaminated data
distributed as $F = F_\rho$\, given by 
\eqref{eq:Frho} where the
true correlation $\rho$ ranges over 
$\{0, 0.05, 0.10, \ldots, 0.95\}$. 
For each $\rho$ we generate $m = 5000$
bivariate data sets $\bZ^j$ with sample 
size $n = 100$.
(We also generated data with $n=20$ 
 yielding the same qualitative
 conclusions.)
We then estimate the bias and the mean 
squared error (MSE) of each correlation
measure $R$ by
\begin{equation}
  \bias_{\rho}(R) = \ave_{j=1}^{m}
	\left(R(\bZ^j)-\rho \right)
	\;\;\; \mbox{ and } \;\;\;
	\MSE_{\rho}(R) = \ave_{j=1}^{m}
  \left( {R(\bZ^j)}-\rho \right)^2\;\;.
\end{equation}

\begin{figure}[!ht]
\centering
\includegraphics[width=0.49\textwidth]
                {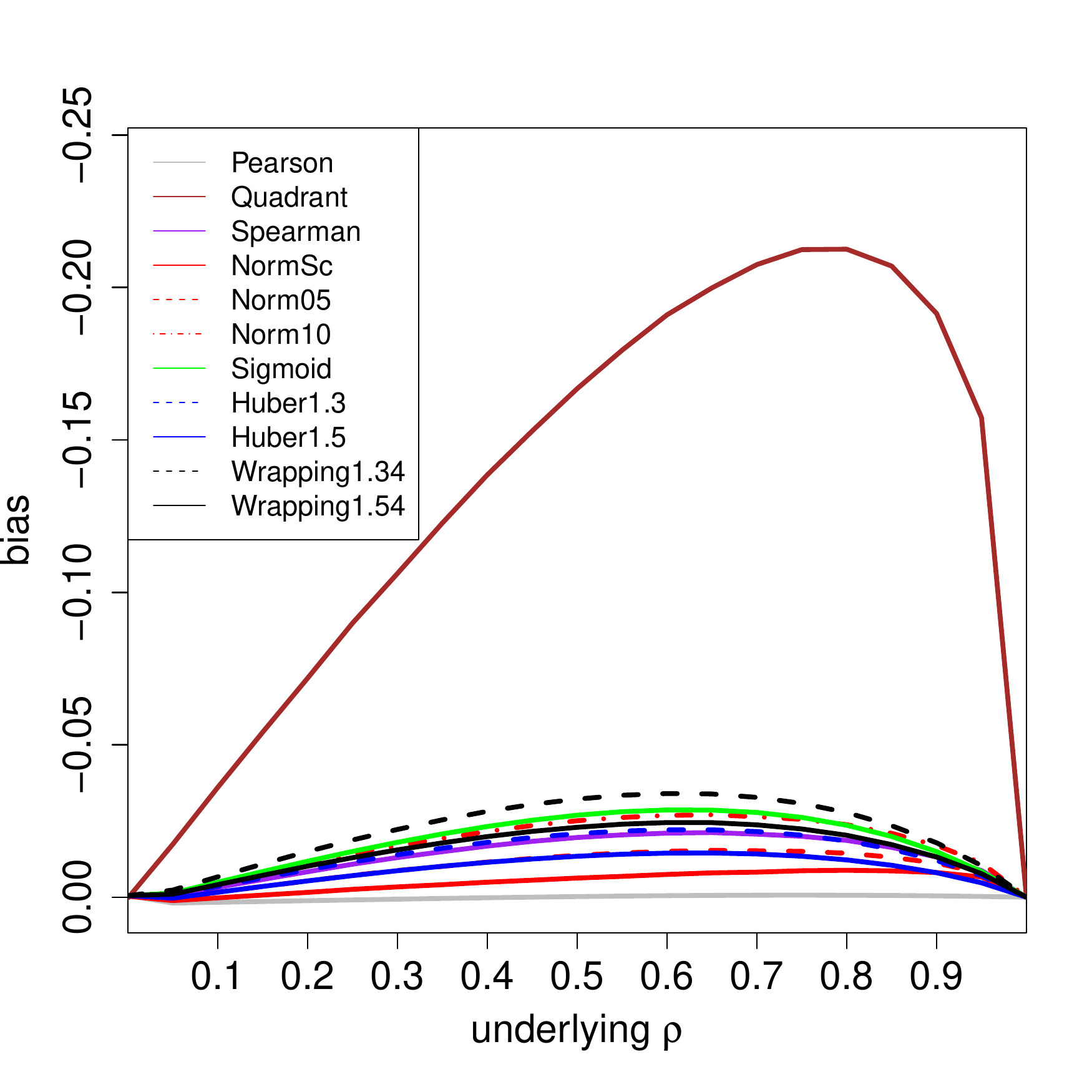}
\includegraphics[width=0.49\textwidth]
                {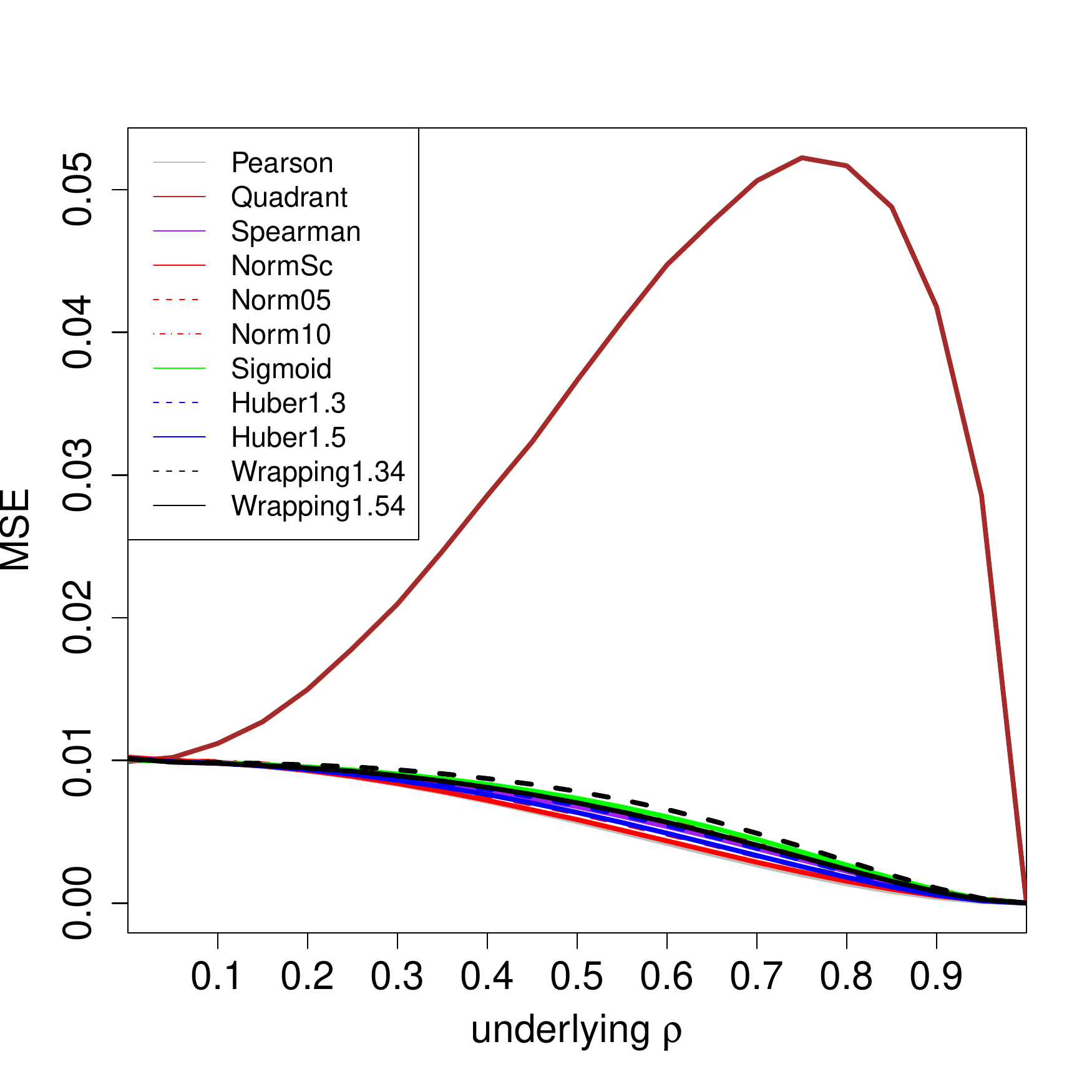}
\vskip-0.2cm						
\caption{Bias and MSE of correlation 
  measures based on transformation, for
	uncontaminated Gaussian data with 
	sample size 100.}
	\label{fig:CBTclean}	
\end{figure}

The bias is shown in the left part of
Figure \ref{fig:CBTclean}.
The vertical axis has flipped signs
because the bias was always negative, 
so $\rho$ is typically underestimated.
Unsurprisingly, the Pearson correlation 
has the smallest bias (known not to be 
exactly zero).
The normal scores correlation and the
Huber $\psi$ with $b=1.5$ are fairly 
close, followed by truncated normal
scores, Spearman and the sigmoid.
Wrapping with $b=1.5$ and $b=1.3$
(both with $c=4$) comes next, still 
with a fairly small bias.
The bias of the quadrant correlation
is much higher.
Note that we could have reduced the bias
of all of these methods by applying the
consistency function $\xi^{-1}$ of
\eqref{eq:xi}, which can be computed
numerically. 
But such consistency corrections
would destroy the crucial PSD property 
for the higher-dimensional data that  
motivate the present work, so we will
not use them here.

The right panel of Figure 
\ref{fig:CBTclean} shows the MSE of 
the same methods, with a pattern
similar to that of the bias.
Even for $n=20$ the bias dominated
the variance (not shown).

{\bf Contaminated data.}
In order to compare the robustness of 
these correlation measures we now add
outliers to the data.
Since the true correlation $\rho$
ranges over positive values here, we will 
try to bring the correlation measures down.
From the proof of Proposition 
\ref{prop:corbias} in 
Section \ref{A:proofbias}
we know that the outliers have the biggest
downward effect when placed at points
$(k,-k)$ and $(-k,k)$ for some $k$.
Therefore we will generate outliers from
the distribution 
\begin{equation*}
H\; = \;\frac{1}{2} N\left(
\begin{bmatrix}
k \\
-k\\
\end{bmatrix}
,0.01^2I\right) + \frac{1}{2}
 N\left(
\begin{bmatrix}
-k \\
k\\
\end{bmatrix}
,0.01^2I\right)
\end{equation*}
for different values of $k$.
The simulations were carried out for 10\%, 
20\% and 30\% of outliers, but we only 
show the results for 10\% as the
relative performance of the methods did
not change much for the higher 
contamination levels.

\begin{figure}[!ht]
\centering
\includegraphics[width=0.49\textwidth]
         {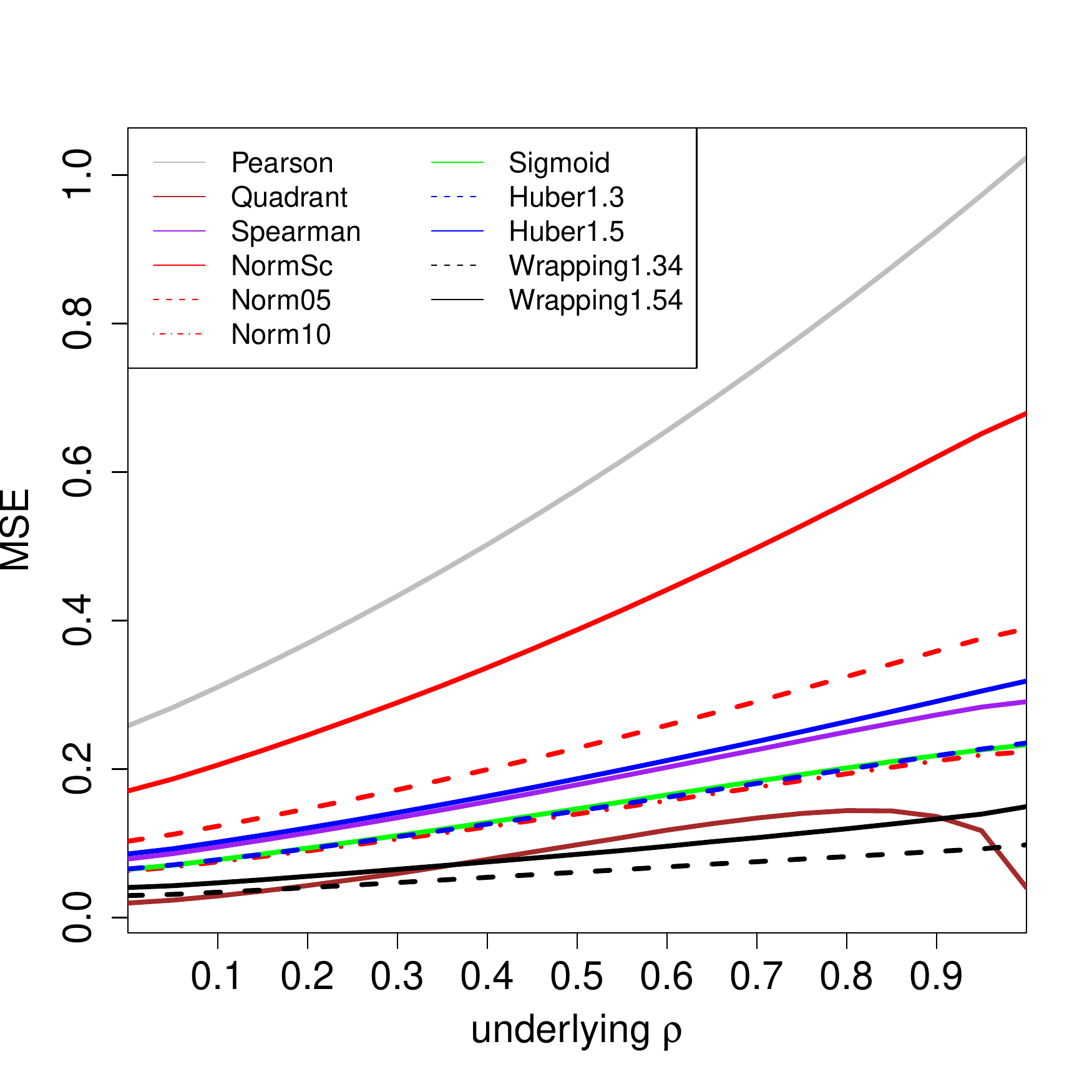}
\includegraphics[width=0.49\textwidth]
         {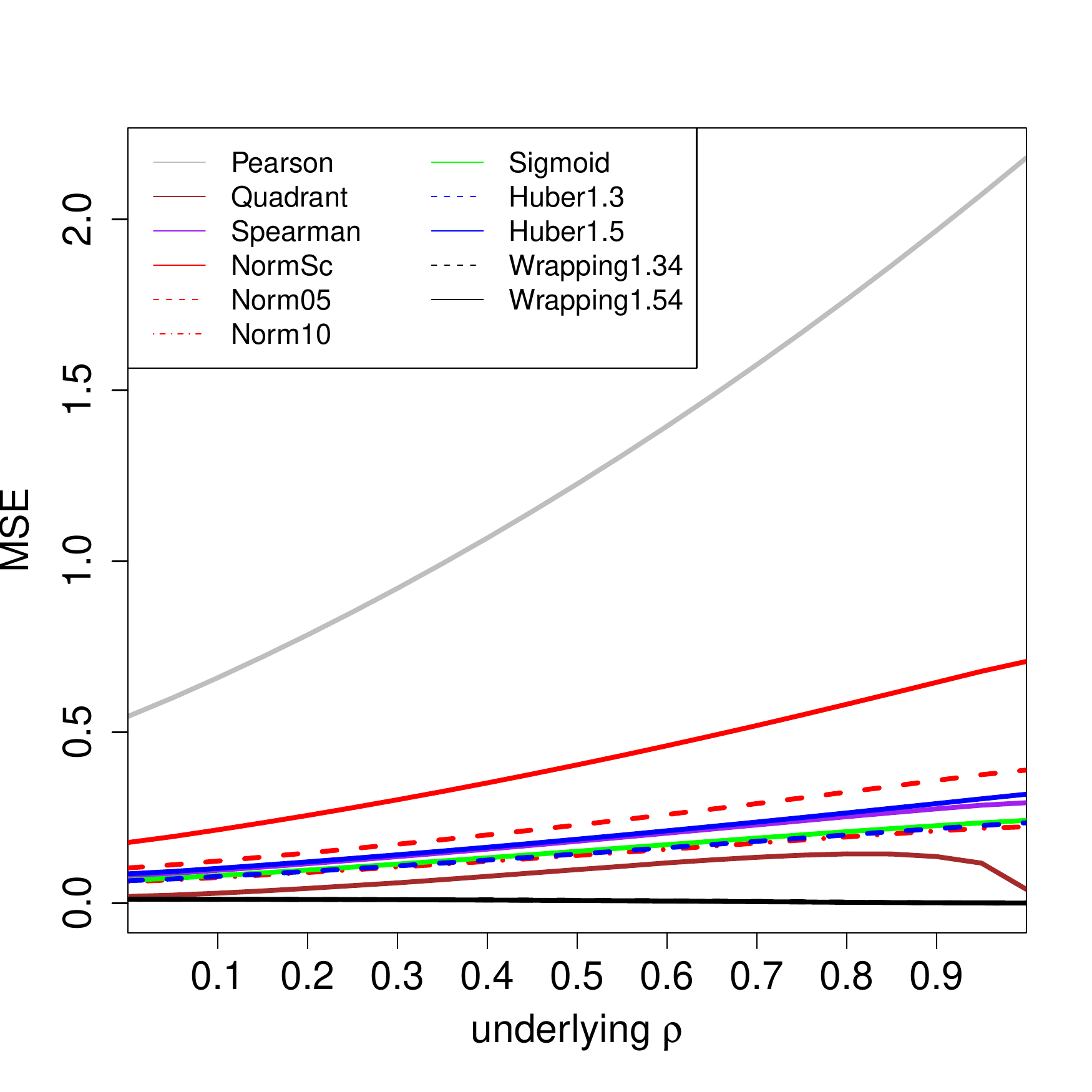}
\vskip-0.2cm						
\caption{MSE of the correlation 
  measures in Figure \ref{fig:CBTclean}
	with 10\% of outliers placed at
	$k = 3$ (left) and $k=5$ (right).}
	\label{fig:CBTcont}	
\end{figure}

The results are shown in Figure 
\ref{fig:CBTcont}
for $k=3$ and $k=5$. 
For $k=3$ we see that the Pearson 
correlation
has by far the highest MSE, followed
by normal scores (whose breakdown
value of 12.4\% is not much higher
than the 10\% of contamination).
The 5\% truncated normal scores and 
the Huber with $b=1.5$ do better,
followed by the Spearman, the sigmoid,
the 10\% truncated normal scores
and the Huber with $b=1.3$.
The quadrant correlation does best
among all the methods based on a
monotone transformation.
However, wrapping still outperforms
it, because it gives the outliers
a smaller weight.
Even though wrapping has a slightly 
lower efficiency for clean data than
Huber's $\psi_b$ with the same $b$, 
in return it delivers more 
resistance to outliers further 
away from the center.

For $k=5$ the pattern is the same,
except that the Pearson correlation
is affected even more and wrapping
has given a near-zero weight 
to the outliers. For $k=2$ (not 
shown) the contamination is not really 
outlying and all methods performed
about the same, whereas for $k > 5$
the curves of the non-Pearson
correlations remain as they are
for $k=5$ since all of our 
transformations $g$ are constant in 
that region.

{\bf Comparison with other 
robust correlation methods.}
As described in the introduction, 
several good robust alternatives to the 
Pearson correlation exist that do not fall 
in our framework.
We would like to find out how well wrapping 
stacks up against the most well-known
of them, such as Kendall's tau.
We also compare with the 
Gnanadesikan-Kettenring (GK) approach
\eqref{eq:GK} in which we replace the 
variance by the square of a robust scale,
in particular the MAD and the scale
estimator $Q_n$ of \cite{Rousseeuw:scale}.

For the approach starting with the
estimation of a robust covariance matrix
we consider the Minimum Covariance
Determinant (MCD) method 
\citep{Rousseeuw:MultBD} using the algorithm
in \citep{Hubert:DetMCD}, and the
Spatial Sign Covariance Matrix (SSCM) of
\cite{Visuri:Rank}.
In both cases we compute a correlation
measure between variables $X_1$ and $X_2$ 
from the estimated scatter matrix $C$ by
\eqref{eq:Cov2Cor}.
For our bivariate generated data the 
matrix $C$ is only $2 \times 2$, 
but if the original data have more 
dimensions the estimated correlation
between $X_1$ and $X_2$ now also depends
on the other variables.
To illustrate this we computed the MCD 
and the SSCM also in $d=10$ dimensions 
where the true covariance matrix is 
given by 
$\Sigma_{jk} = \rho$ for $j \neq k$
and 1 otherwise. The simulation then
reports the result of \eqref{eq:Cov2Cor}
on the first two variables only.

\begin{figure}[!ht]
\centering
\includegraphics[width=0.49\textwidth]
                {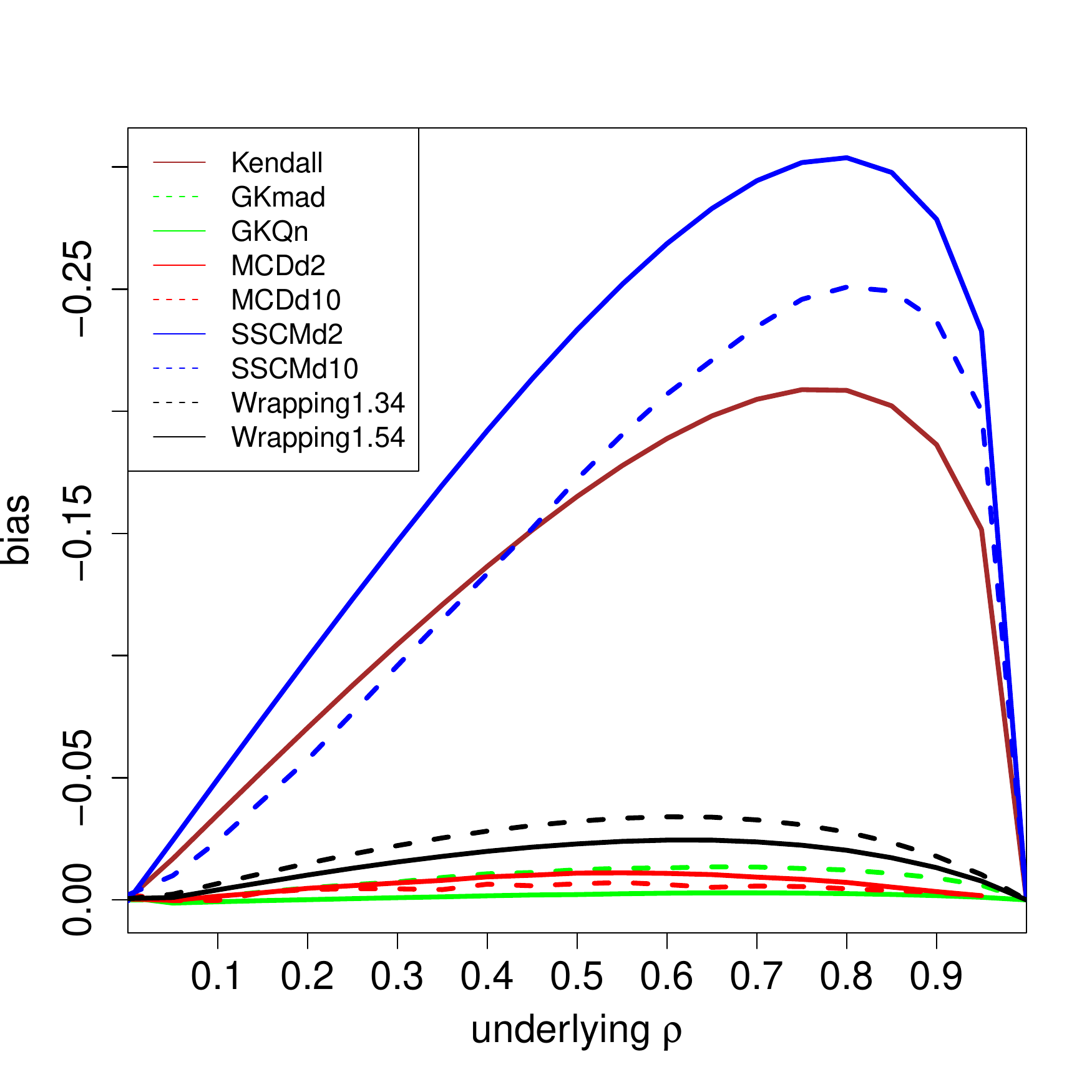}
\includegraphics[width=0.49\textwidth]
                {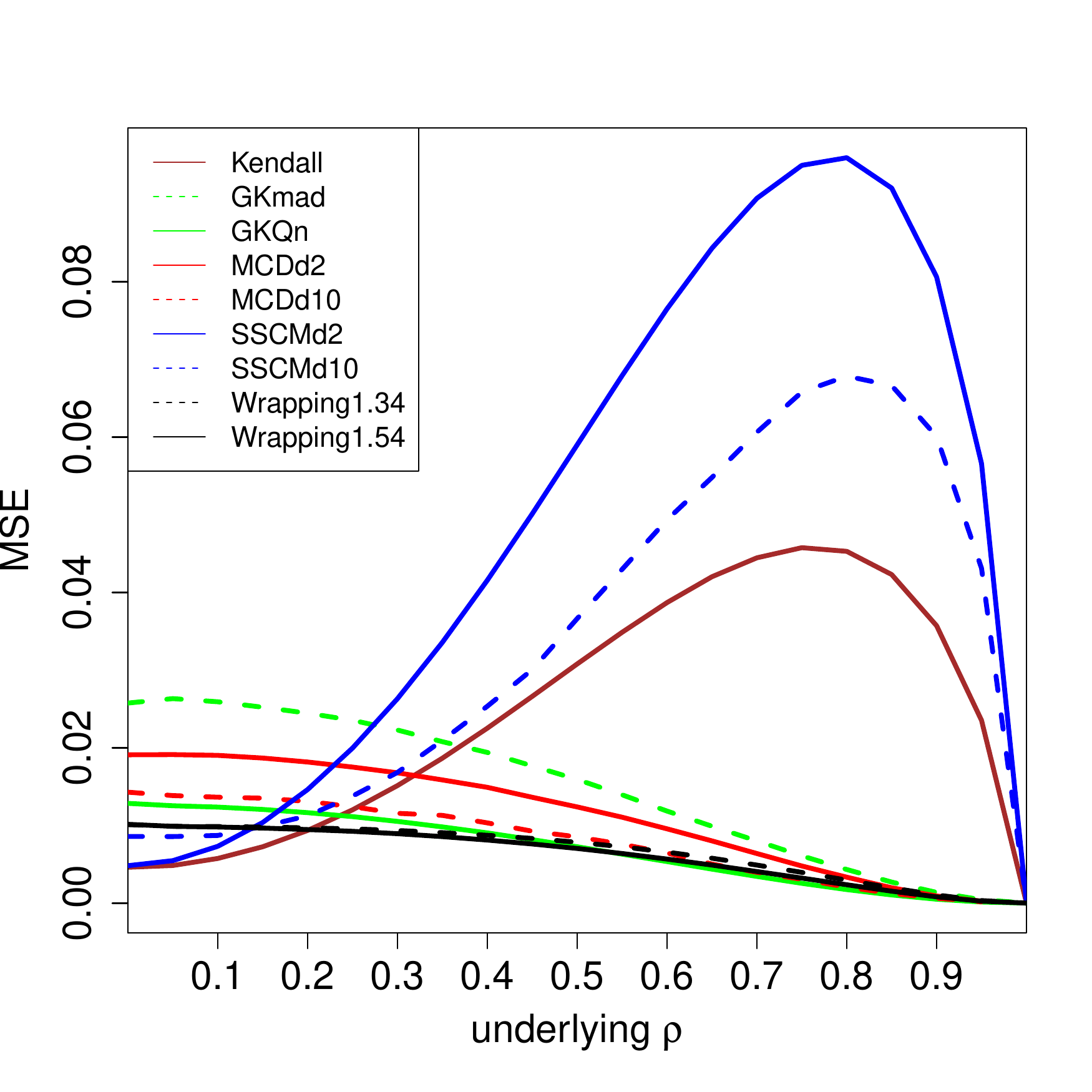}
\vskip-0.2cm						
\caption{Bias and MSE of other robust
  correlation measures, for
	uncontaminated Gaussian data with 
	sample size 100.}
	\label{fig:CCclean}	
\end{figure}

The left panel of Figure \ref{fig:CCclean} 
shows the bias of all these methods, in 
the same setting as Figure
\ref{fig:CBTclean}.
The two GK methods and the MCD 
computed in 2 and 10 dimensions have the 
smallest bias, followed by wrapping.
The Kendall bias is substantially larger, 
and in fact looks similar to the bias of 
the quadrant correlation in Figure 
\ref{fig:CCclean}, which is not so
surprising since they possess the same
function 
$\xi(\rho) = 2 \arcsin(\rho)/\pi$
in \eqref{eq:xi}.
The bias of the SSCM is even larger,
both when computed in $d=2$ dimensions
and in $d=10$.
The MSE in the right panel of 
Figure \ref{fig:CCclean} shows a
similar pattern.

\begin{figure}[!ht]
\centering
\includegraphics[width=0.49\textwidth]
       {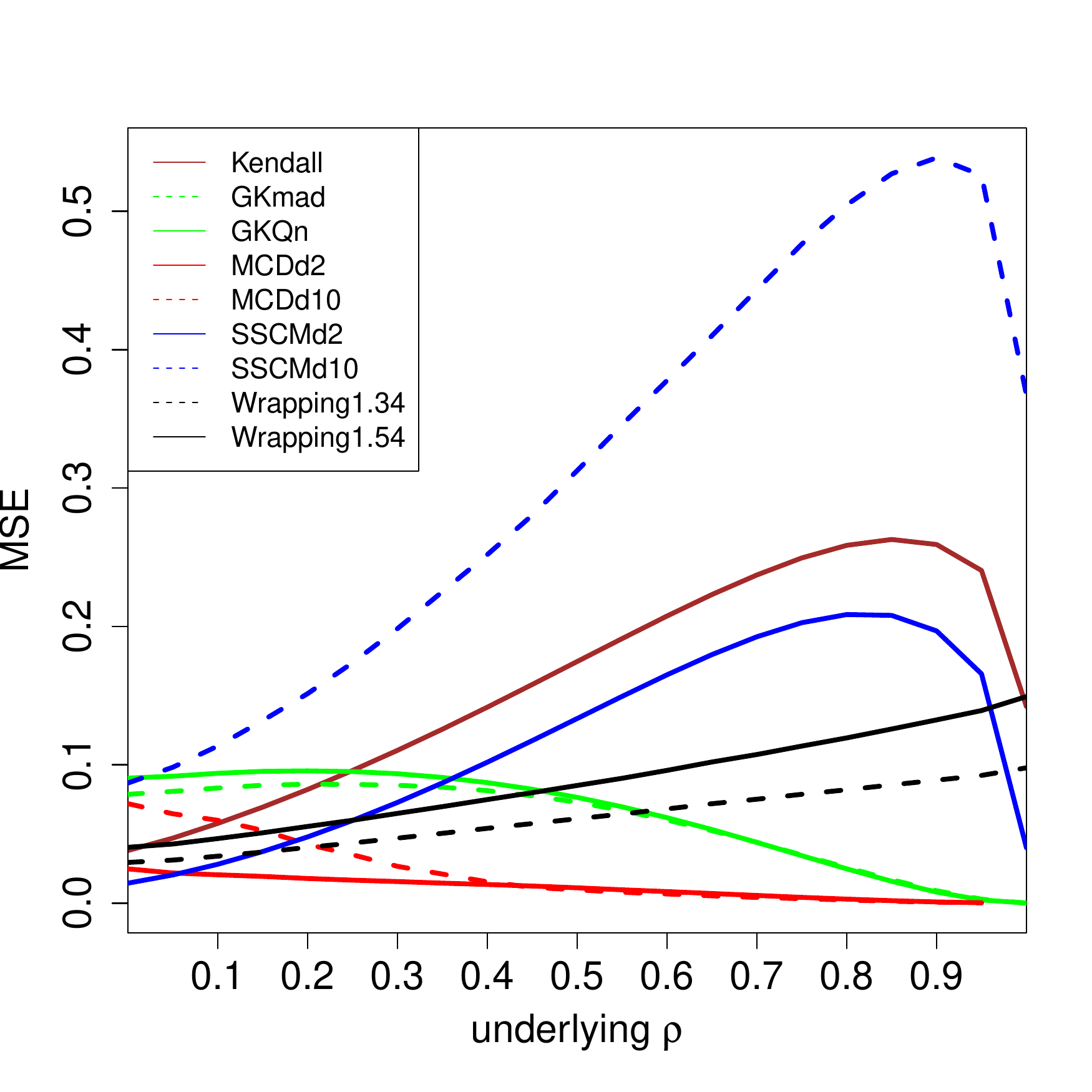}
\includegraphics[width=0.49\textwidth]
       {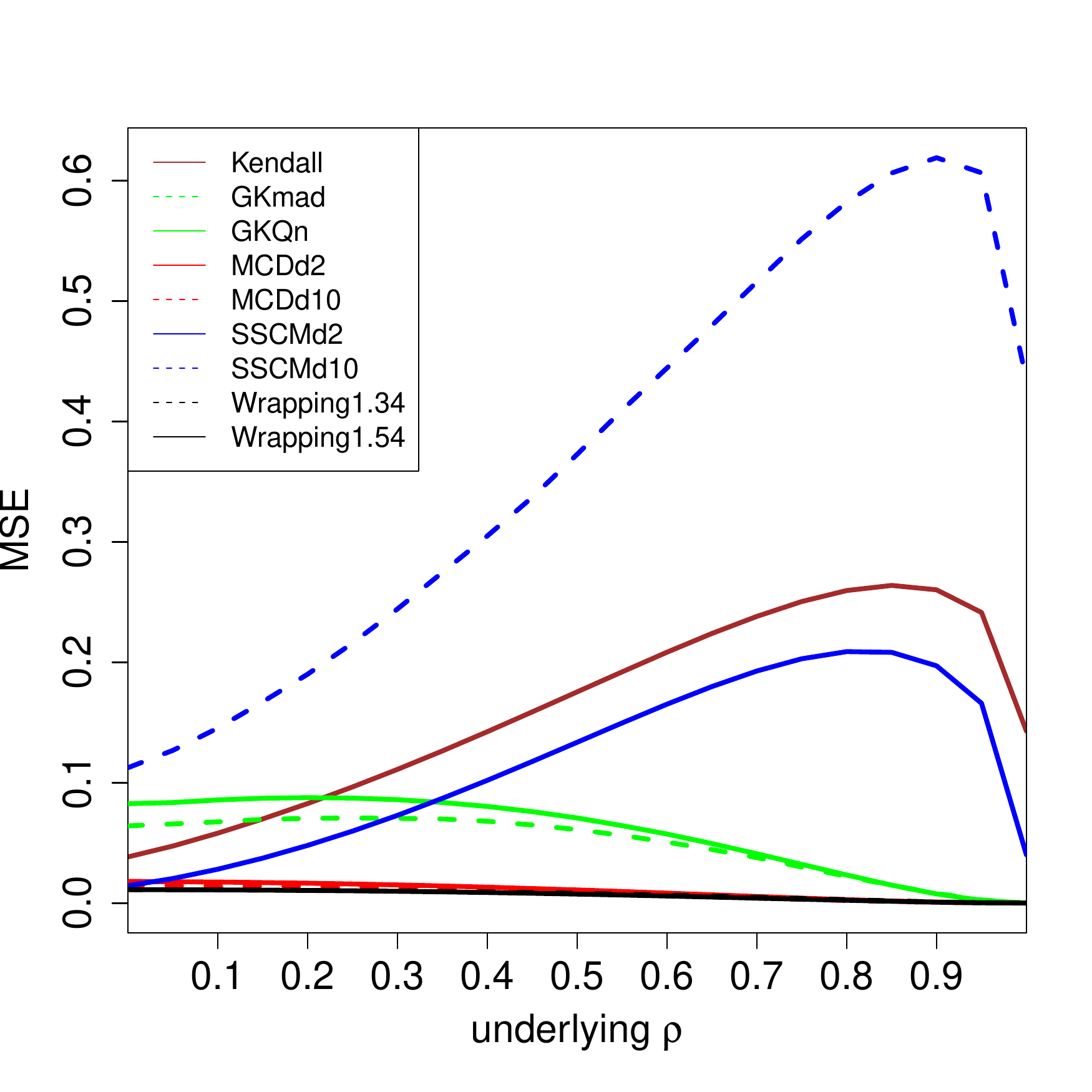}
\vskip-0.2cm						
\caption{MSE of the correlation 
  measures in Figure \ref{fig:CCclean}
	with 10\% of outliers placed at
	$k = 3$ (left) and $k=5$ (right).}
	\label{fig:CCcont}	
\end{figure}

Figure \ref{fig:CCcont} shows the effect
of 10\% of outliers, using the same 
generated data as in
Figure \ref{fig:CBTcont}.
The left panel is for $k=3$. 
The scale of the vertical axis indicates
that the outliers have increased the MSE
of all methods.
The MCD in $d=2$ dimensions is the least
affected, whereas 
the GK methods, the SSCM with $d=2$ and 
Kendall's tau are more sensitive.
Note that the data in $d=10$ dimensions 
was only contaminated in the first 2 
dimensions, and the MCD still does quite 
well in that setting.
On the other hand, the MSE of the SSCM 
in $d=10$ is now much higher. 

To conclude, wrapping holds its own even
among well-known robust correlation 
measures outside our transformation 
approach.
Wrapping was not the overall best method
in our simulation, that would be the MCD, 
but the latter requires much more 
computation time which goes up a lot in 
high dimensions.
Moreover, the highly robust quadrant 
transformation yields a low efficiency
as it ignores much information 
in the data.

Therefore, wrapping seems a good choice
for our purpose, which is to construct 
a fast robust method for fitting 
high dimensional data.
Some other methods like the MCD perform 
better in low dimensions (say, upto 20), 
but in high dimensions the MCD and related
methods become infeasible, whereas the
SSCM does not perform well any more.

\section{Use in higher dimensions}
\label{sec:highdim}

\subsection{Methodology}
\label{sec:method}
So far the illustrations of wrapping were
in the context of bivariate correlation.
In this section we explain its
use in the higher-dimensional context
for which it was developed.
Our approach is basically to wrap
the data first, carry out an existing
estimation technique on the wrapped 
data, and then use that fit for the 
original data.
We proceed along the following steps.

{\bf Step 1: estimation.} 
For each of the (possibly many)
continuous variables $X_j$ with
$j=1,\ldots,d$ we compute a robust
initial scale estimate $\hs_j$
such as the MAD.
Then we compute a one-step location
M-estimator $\hmu_j$ with the
wrapping function $\psi_{b,c}$
with defaults $b=1.5$ and $c=4$.
We could take more steps or 
iterate to convergence, but this 
would lead to a higher contamination
bias \citep{Rousseeuw:kStepM}.

{\bf Step 2: transformation.} 
Next we wrap the continuous variables.
That is, we transform any $x_{ij}$ to
\begin{equation}\label{eq:wrapx}
  x_{ij}^* \;=\; g(x_{ij}) \;=\;
	 \hmu_j + \hs_j \, \psi_{b,c}
	 \Big(\frac{x_{ij}-\hmu_j}
	           {\hs_j}\Big)\;\;.
\end{equation}
Note that $\ave_i(\tx_{ij})$ is a robust
estimate of $\mu_j$  and
$\std_i(\tx_{ij})$ is a robust
estimate of $\sigma_j$\,.
The wrapped variables $X^*_j$ do not
contain outliers, and when the original
$X_j$ is Gaussian over 86\% of its
values remain unchanged, that is
$\tx_{ij} = x_{ij}$\;.
If $x_{ij}$ is missing we have to assign a
value to $g(x_{ij})$ in order to preserve 
the PSD property of product moment matrices,
and $g(x_{ij})=\hat{\mu}_j$ is the natural 
choice.
We do not transform discrete variables -- 
depending on the context one may or may 
not leave them out of the subsequent 
analysis.

{\bf Step 3: fitting.} 
We then fit the wrapped data $\tx_{ij}$ by 
an existing multivariate method, yielding 
for instance a covariance matrix
or sparse loading vectors. 

{\bf Step 4: using the fit.} 
To evaluate the fit we will look at the
deviations (e.g. Mahalanobis distances) 
of the wrapped cases $\bx^*_i$ as well as
the original cases $\bx_i$\,. 

Note that the time complexity of Steps 1 
and 2 for all $d$ variables is only $O(nd)$.
Any fitting method in Step 3
must read the data so its complexity is 
at least $O(nd)$.
Therefore the total 
complexity is not increased	by wrapping,
as illustrated in Table \ref{tab:times}.

\subsection{Estimating covariance and
precision matrices}
\label{sec:specific}

{\bf Covariance matrices.}
The covariance matrix of the wrapped 
variables has the entries 
\begin{equation}\label{eq:Cjk}
  C(j,k) = \Cov(X^*_j,X^*_k) =
	\hs_j \, \hs_k \,
	\Cor\big( \psi_{b,c}
	 \Big(\frac{x_{ij}-\hmu_j}
	           {\hs_j}\Big),
	\psi_{b,c}
	 \Big(\frac{y_{ik}-\hmu_k}
	           {\hs_k}\Big) \big) \;.
\end{equation}
for $j,k = 1,\ldots,d$. The resulting 
matrix is clearly PSD.
We also have the independence property:
if variables $X_j$ and $X_k$ are 
independent so are $X^*_j = g(X_j)$
and $X^*_k = g(X_k)$, and as these
are bounded their population covariance 
exists and is zero.
 
\cite{Oellerer:robprec} defined robust
covariances with a formula like 
\eqref{eq:Cjk} in which the correlation 
on the right was a rank correlation.
They showed that the explosion breakdown 
value of the resulting scatter matrix 
(i.e. the percentage of outliers
required to make its largest eigenvalue
arbitrarily high) is at least that of
the univariate scale estimator $S$ 
yielding $\hs_j$ and $\hs_k\,$, and their 
proof goes through without changes in 
our setting.
Therefore, the robust covariance matrix 
\eqref{eq:Cjk} also has an
explosion breakdown value of 50\%. 

The scatter matrix given by 
\eqref{eq:Cjk} is easy to compute, 
and can for instance be used for 
anomaly detection.
In Section \ref{A:robdist} of the 
Supplementary Material it is illustrated how 
robust Mahalanobis distances obtained from 
the estimated scatter matrix can detect 
outlying cases.
The scatter matrix can also be used
in other multivariate methods such as 
canonical correlation analysis, and
serve as a fast initial estimate in the 
computation of other robust
methods such as \citep{Hubert:DetMCD}.

{\bf Precision matrices
     and graphical models.}
The precision matrix is the inverse of
the covariance matrix, and allows to
construct a Gaussian graphical
model of the variables.
\cite{Oellerer:robprec} and 
\cite{Tarr:robprec} estimated
the covariance matrix from rank 
correlations, but one could also
use wrapping for this step.
When the dimension $d$ is too high 
the estimated covariance matrix 
cannot be inverted, so
these authors construct a sparse
precision matrix by applying 
GLASSO.
\cite{Oellerer:robprec} show that
the breakdown value of the resulting
precision matrix, for both implosion
and explosion, is as high as that
of the univariate scale estimator.
This remains true for wrapping,
so the resulting robust precision
matrix has breakdown value 50\%. 

\subsection{Distance Correlation}
\label{sec:dependence}
There exist measures of dependence 
which do not give rise to PSD matrices but 
are used as test statistics for dependence, 
such as mutual information and the distance 
correlation of \cite{Szekely:distcor}, 
which yield a single nonnegative scalar
that does not reflect the direction of the
relation if there is one.
The theory of distance correlation only
requires the existence of first moments.
The distance correlation $\mbox{dCor}$
between random vectors $\bX$ and $\bY$ is defined 
through the Pearson correlation between the doubly 
centered interpoint distances of $\bX$ and
those of $\bY$.
It always lies between 0 and 1.
The population version $\mbox{dCor}(\bX,\bY)$
can be written in terms of the characteristic 
functions of the joint distribution of
$(\bX,\bY)$ and the marginal distributions of
$\bX$ and $\bY$. This allows 
\cite{Szekely:distcor} to prove that
$\mbox{dCor}(\bX,\bY)=0$ implies that $\bX$ 
and $\bY$ are independent, a property that
does not hold for the plain Pearson 
correlation.

The population $\mbox{dCor}(\bX,\bY)$ 
is estimated by its finite-sample version
$\mbox{dCor}(\bX_n,\bY_n)$ which is used as
a test statistic for dependence.
For a sample of size $n$ this 
would appear to require $O(n^2)$ computation 
time, but there exists an $O(n\log(n))$ 
algorithm \citep{Huo:fast} for
the bivariate setting.

\begin{figure}[!t]
\centering
\vskip0.5cm
\includegraphics[width=0.99\textwidth]
 								{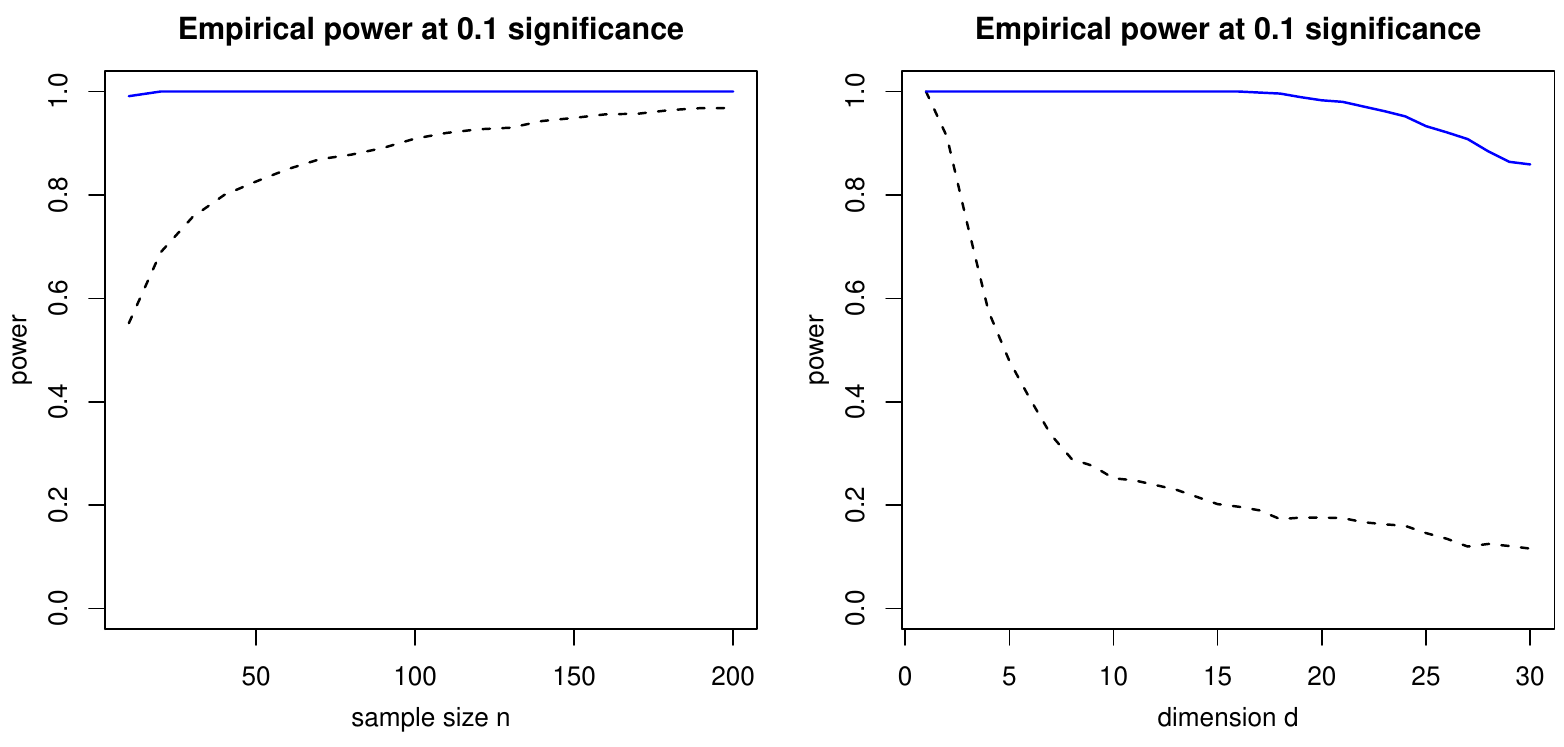}								
\vskip-0.2cm						
\caption{Left panel: power of dCor 
(dashed black curve)
and its robust version (blue curve) 
for bivariate 
$\bX$ and $\bY$ with distribution $t(1)$ and
independence except for $\bX_1=\bY_1$ versus the 
sample size $n$.
Right panel: power of dCor and its robust version
for $d$-dimensional $\bX$ and $\bY$ with 
distribution $t(1)$ and $n=100$,
as a function of the dimension $d$.}
\label{fig:distcor_Cauchy}	
\end{figure}

By itself distance correlation
is not robust to outliers in the data.
In fact, we illustrate in Section \ref{A:distcov}
of the Supplementary Material that the distance
correlation of independent variables can be 
made to approach 1 by a single outlier among
$100,000$ data points, and the distance 
correlation of perfectly dependent 
variables can be made to approach zero.
On the other hand, we could first transform the
data by the function $g$ of \eqref{eq:wrapx} 
with the sigmoid $\psi(z) = \tanh(z)$,
and then compute the distance covariance.
This combined method does not require the first
moments of the original variables to exist,
and the population version is again zero if
and only if the original variables are
independent (since $g$ is invertible).
Figure \ref{fig:distcor_Cauchy} illustrates 
the robustness of this combined statistic.

The data for Figure \ref{fig:distcor_Cauchy}
were generated following Example 1(b) in 
\citep{Szekely:distcor},  
where $\bX$ and $\bY$ are 
multivariate and all their components follow 
$t(1)$, the Student $t$-distribution with 
one degree of freedom.
The null hypothesis states that $\bX$ and 
$\bY$ are independent. We investigate the 
power of the test for dependence under the
alternative that all components of $\bX$
and $\bY$ are independent except for 
$\bX_1 = \bY_1$. For this we use the
permutation test implemented as 
{\it dcor.test} in the R package {\it energy}.
As in \citep{Szekely:distcor} we set the 
significance level to 0.1. The empirical
power of the test is then the fraction of
the $1000$ replications in which the test
rejects the null hypothesis.

In the left panel of Figure
\ref{fig:distcor_Cauchy} we see the empirical
power as a function of the sample size when
$\bX$ and $\bY$ are both bivariate.
The power of the original dCor (dashed black 
curve) starts around 0.6 for $n=20$ and 
approaches 1 when $n = 200$. 
This indicates that for small sample sizes
the components $\bX_2$ and $\bY_2$, even though 
they are independent of everything else, have 
added noise in the doubly centered distances.
In contrast, the power of the robust method 
(solid blue curve) is close to 1 overall.
No outliers were added to the data, but the
underlying distribution t(1) is long-tailed.

The right panel of Figure
\ref{fig:distcor_Cauchy} shows the effect of
increasing the dimension $d$ of $\bX$ and $\bY$,
for fixed $n=100$.
At dimension $d=1$ we only have the components
$\bX_1=\bY_1$ and both methods have power 1.
At dimension $d=2$, 
dCor has power 0.9 and the
robust version has power 1. When increasing
the dimension further, the power of dCor
goes down to about 0.3 around dimension $d=8$,
whereas the power of the robust method 
only starts going down around dimension 
$d=17$ and is still reasonable at dimension 
$d=30$.
This illustrates that the transformation has
tempered the effect of the $d-1$
independent variables on the doubly centered
distances, delaying the curse of 
dimensionality in this setting.

\subsection{Fast detection of anomalous cells}
\label{sec:FastDDC}

Wrapping is a coordinatewise approach
which makes it especially robust against
cellwise outliers, that is, anomalous
cells $x_{ij}$ in the data matrix.
In this paradigm a few cells in a row
(case) can be anomalous whereas many
other cells in the same row still 
contain useful information, and in such
situations we would rather not remove 
or downweight the entire row.
The cellwise framework was first 
proposed and studied by
\cite{Alqallaf:scalable,Alqallaf:Propag}.

Most robust techniques developed in the
literature aim to protect against
rowwise outliers.
Such methods tend not to work well in 
the presence of cellwise
outliers, because even a relatively
small percentage of outlying cells
may affect a large percentage of the
rows.
For this reason several authors have 
started to develop cellwise robust 
methods
\citep{Agostinelli:cellwise}. 
In the bivariate simulation of 
Section \ref{sec:sim} we generated
rowwise outliers, but the results 
for cellwise outliers are similar  
(see Section \ref{A:cell}
in the Supplementary Material).

Actually {\it detecting} outlying cells 
in data with many dimensions is not 
trivial, because the correlation between 
the variables plays a role.
The DetectDeviatingCells (DDC) method 
of \cite{Rousseeuw:DDC}
predicts the value of each cell from
the columns strongly correlated with
that cell's column.
The original implementation of DDC
required computing all $O(d^2)$ 
robust correlations between the 
$d$ variables, yielding total time
complexity $O(nd^2)$ which grows fast 
in high dimensions.

Fortunately, the computation time can
be reduced a lot by the wrapping method.
This is because the product moment 
technology allows for nice shortcuts. 
Let us standardize two 
column vectors (that is, variables) 
$X_n = (x_1,\ldots,x_n)^T$ and $Y_n$
to zero mean and unit standard deviation. 
Then it is easy to verify that their
correlation satisfies
\begin{equation}\label{eq:cordist}
  \Cor(X_n,Y_n) \;=\; \frac{1}{n-1}
  \big\langle X_n,Y_n \big\rangle \;=\; 
	1 - \frac{||X_n - Y_n||^2}{2(n-1)}
\end{equation}
where $||\ldots||$ is the usual
Euclidean distance.
This monotone decreasing relation between
correlation and distance allows us to
switch from looking for high 
correlations in $d$ dimensions to
looking for small distances in $n$
dimensions. When $n << d$ this is very 
helpful, and used e.g. in Google 
Correlate \citep{Vanderkam:Google}.

The identity \eqref{eq:cordist} can be
exploited for robust correlation by
wrapping the variables first.
In the (ultra)high dimensional case we
can thus transpose our dataset so it
becomes $d \times n$.
If needed we can reduce its dimension
even more to some $q << n$ by computing 
the main principal components and 
projecting on them, 
which preserves the Euclidean 
distances to a large extent.

Finding the $k$ variables 
that are most correlated to a 
variable $X_j$\; therefore 
comes down to finding its $k$ nearest 
neighbors in $q$-dimensional space.
Fortunately there exist fast 
approximate nearest neighbor 
algorithms \citep{Arya:ANN} that can 
obtain the $k$ nearest neighbors of 
all $d$ points in $q$ dimensions in
$O(qd\log(d))$ time, a big
improvement over $O(nd^2)$.
Note that we want to find both large 
positive and large 
negative correlations, so we look 
for the $k$ nearest neighbors in
the set of all variables and 
their sign-flipped versions.

Using these shortcuts we constructed
the method FastDDC which takes far
less time than the original DDC and can 
therefore be applied to data in much 
higher dimensions.
The detection of anomalous cells 
will be illustrated in the real data 
examples in Section \ref{sec:app}.
In both applications, finding the 
anomalies is the main result of the 
analysis.

\section{Real data examples}
\label{sec:app}

\subsection{Prostate data}
\label{sec:prostate}

In a seminal paper, \cite{Singh:prostate} 
investigated the
prediction of two different types of 
prostate cancer from genomic information.
The data is available
as the R file Singh.rda in\\
{\it http://www.stats.uwo.ca/faculty/aim/2015/9850/microarrays/FitMArray/data/}
and contains 12600 genes. 
The training set consists of 102 patients
and the test set has 34.
There is also a response variable with
the clinical classification, -1 for
tumor and 1 for nontumor.

\begin{figure}[!ht]
\centering
\includegraphics[width=0.55\textwidth]
   {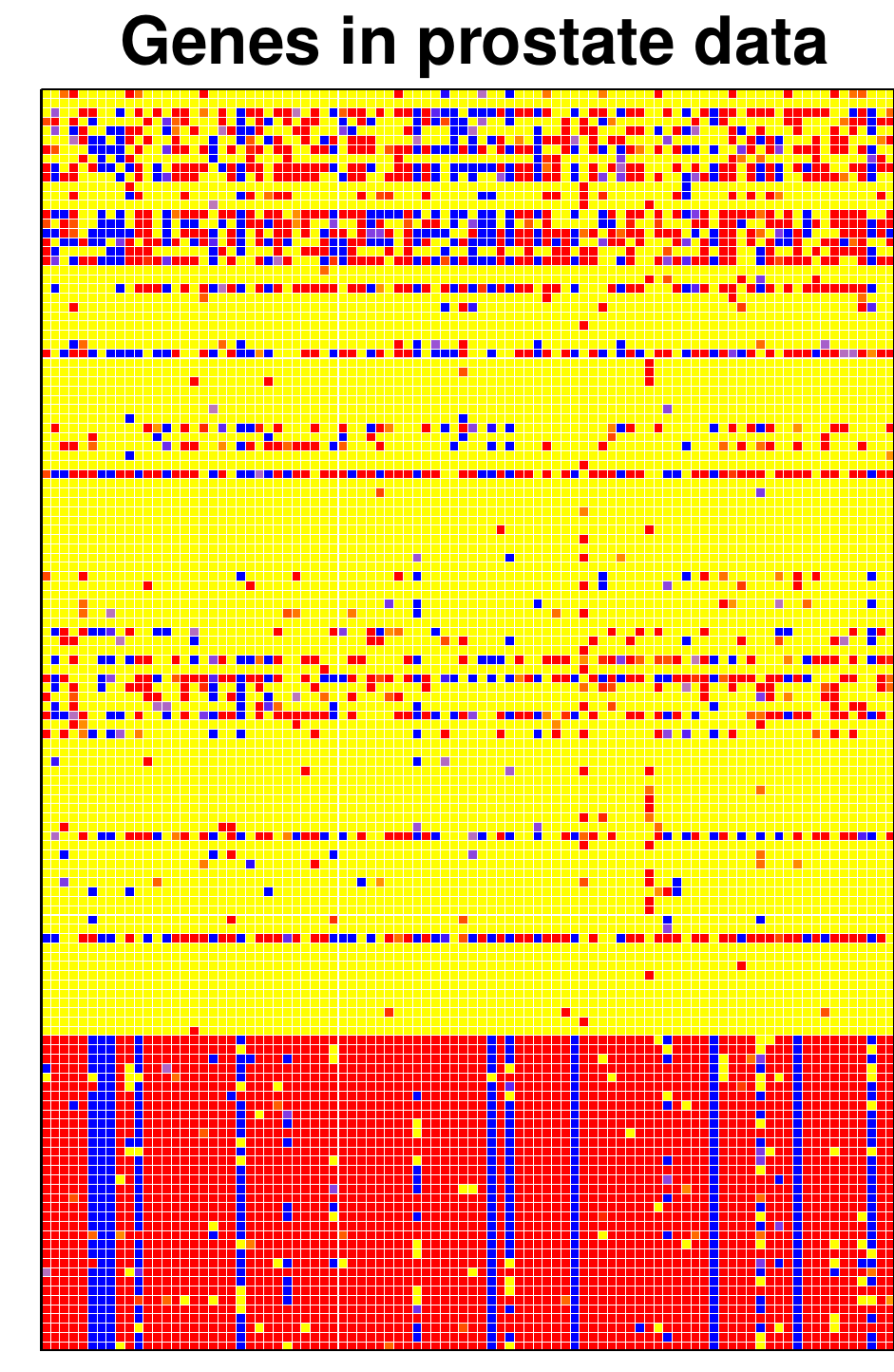}
\vskip-0.2cm						
\caption{Prostate data: cellmap of the 
  genes with the largest number of 
	flagged cells.}
	\label{fig:prostate}	
\end{figure}

With the fast version of DDC introduced
in Subsection \ref{sec:FastDDC} we can now
analyze the entire genetic data set with
$n=136$ and $d=12600$, which
would take very long with the 
original DDC algorithm. 
Now it takes under 1 minute on a laptop.
In this analysis only the genetic data 
is used and not the response variable, and 
the DDC method is not told which rows 
correspond to the training set.
Out of the 136 rows 33 are flagged as
outlying, corresponding to the test set
minus one patient.
The entire cellmap of size
$136 \times 12600$ is hard to visualize.
Therefore we select the 100 variables 
with the most flagged cells, yielding the 
cellmap in Figure \ref{fig:prostate}. 
The flagged cells are colored 
red when the observed value 
(the gene expression level)
is higher than predicted, and blue when
it is lower than predicted.
Unflagged cells are colored yellow.

The cellmap clearly shows that the bottom
rows, corresponding to the test set, 
behave quite differently from the others.
Indeed, it turns out that the
test set was obtained by a different 
laboratory. This suggests to align the
genetic data of the test set with that of 
the training set by some form of 
standardization, before applying a model 
fitted on the training data to predict the 
response variable on the test data.

\subsection{Video data}
\label{sec:video}

For our second example we analyze a video 
of a parking lot, filmed by a static 
camera. The raw video can be found on 
{\it http://imagelab.ing.unimore.it/visor} 
in the category \textit{Videos for human 
action recognition in videosurveillance}. 
It was originally analyzed 
by \cite{Ballan:Video} using sophisticated
computer vision technology. 
The video is 23 seconds long and 
consists of 230 Red/Green/Blue (RGB) 
frames of 640 by 480 pixels, so each
frame corresponds with 3 matrices of 
size $640 \times 480$.
In the video we see two men coming from
opposite directions, meeting in the
center where they talk, and then
running off one behind the other.
Figure \ref{fig:videosum} shows 3 frames 
from the video. 
The men move through the scene, so they 
can be considered as outliers. Therefore
every frame (case) is contaminated, but 
only in a minority of pixels (cells).

We treat the video as a dataset $\bX$
with 230 row vectors $\bx_i$ of length
$921,600 = 640\cdot 480 \cdot 3$, and
we want to carry out a PCA based on the
robust covariance matrix between the 
$921,600$ variables.
When dealing with datasets this large one 
has to be careful with memory management, 
as a covariance matrix between these 
variables has nearly $10^{12}$ entries 
which is far too many to store in RAM 
memory.
Therefore, we proceed as follows:

\begin{figure}[!htb]
\centering
\vspace{0.3cm}
\includegraphics[width=1.0\textwidth]
     {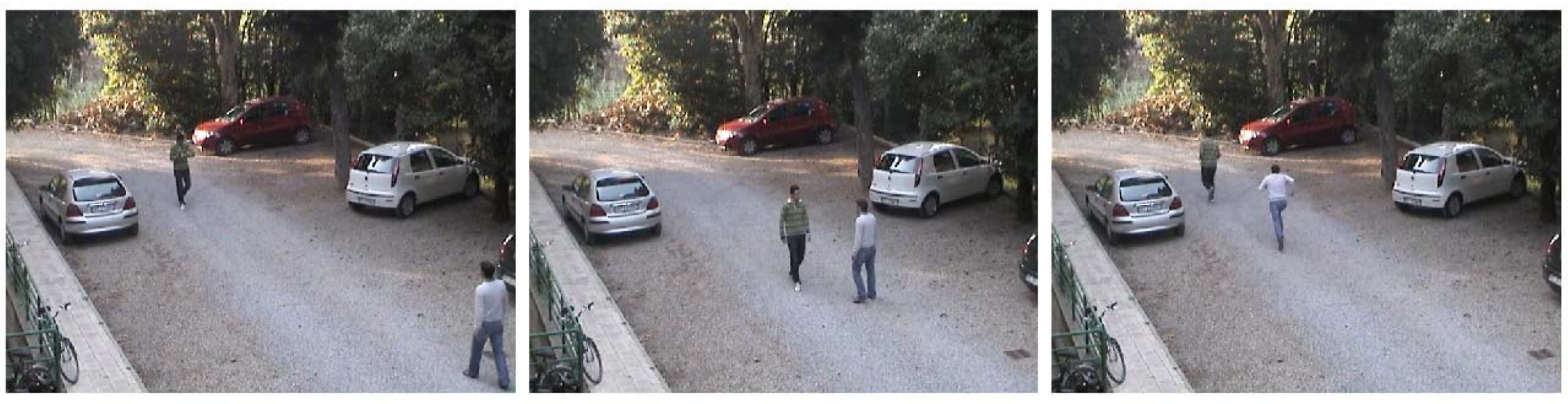}
\caption{Frames 60, 100 and 200 of the
video data.}
\label{fig:videosum}
\end{figure}

\begin{enumerate}
\item Wrap the 230 data values of each RGB 
      pixel (column) $X_j$ which yields the 
			wrapped	data matrix $\bX^*$ and its
			centered version $\bZ^* = 
			\bX^*-\boldsymbol{\overline{x^*}}\;$.		
\item Compute the first $k=3$ loadings of 
      $\Cov(\bX^*) = 
			\frac{n}{n-1}\PM(\bZ^*)$\;.
			We cannot actually compute or store 
			this covariance matrix,
			so instead we perform a truncated 
			singular value decomposition (SVD) of 
			$\bZ^*$	with $k=3$ components, which 
			is mathematically equivalent. 
			For this we use the efficient function 
			{\it propack:svd()} from the R
			package {\it svd} with option
			{\it neig=3}, yielding the loading
			row vectors $\bv_j$ for $j = 1,2,3$.
\item Compute the 3-dimensional robust 
      scores $\bt_i$ by 
      projecting the {\it original} data on 
			the robust loadings obtained from the 
			{\it wrapped} data, i.e. $\bt_i = 
			(\bx_i-\boldsymbol{\overline{x^*}})
			(\bv_1^T,\bv_2^T,\bv_3^T)\,$.
\end{enumerate}

The classical PCA result can be 
obtained by carrying out steps 2 and 3 
on $\bZ = \bX-\boldsymbol{\overline{x}}\;$
without any wrapping. 

We also want to compare with other robust
methods.
For the Spearman method we first replace
each column $X_j$ by its ranks, i.e.
$R_{ij}$ is the rank of $x_{ij}$ among
all $x_{hj}$ with $h=1,\ldots,n$.
We also compute $\hs_j = \MAD(X_j)$.
Then we transform each $x_{ij}$ to
$\,(R_{ij} - \ave_h(R_{hj}))\hs_j/
 \std_h (R_{hj})\,$
yielding a matrix whose columns have 
mean zero and standard deviation $\hs_j$ 
to which we again apply step 2.
Another method  is to transform the data 
as in \eqref{eq:wrapx} but 
using Huber's $\psi$ function
$\psi_b(z) = [z]_{-b}^{b}$
with the same $b=1.5$ as in wrapping.

\begin{figure}[!htb]
\centering
\vspace{0.3cm}
\includegraphics[width=1\textwidth]
		{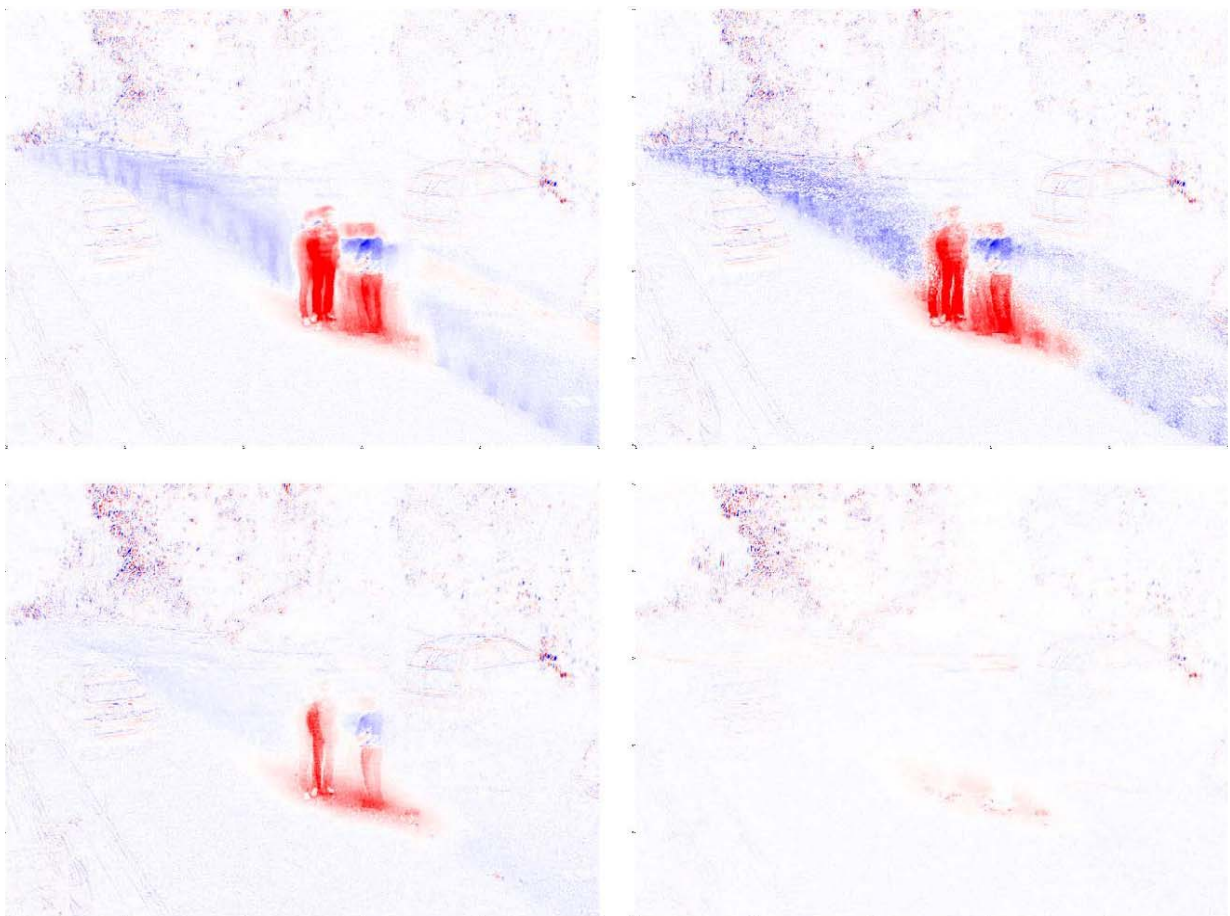} 
\caption{First loading vector of the video 
data, for classical PCA (upper left), 
Spearman correlation (upper right),
Huber's $\psi$ (lower left), and
wrapping (lower right).}
\label{fig:loadings}
\end{figure}

Figure \ref{fig:loadings} shows the first 
loading vector $\bv_1$ displayed as an 
image, for all 4 methods considered.
Positive loadings are shown in red,
negative ones in blue, and loadings near
zero look white.
For wrapping the loadings basically
describe the background, whereas for
classical PCA they are affected by the
moving parts (mainly the men and some 
leaves) that are outliers in this setting.
The Spearman loadings resemble those 
of the classical method, whereas those
with Huber's $\psi$ are in between.
Similar conclusions hold for the second
and third loading vectors (not shown).

We can now compute a fit to each
frame. For wrapping this is 
$\,\boldsymbol{\hat{x_i}} =
\bt_i\, (\bv_1^T,\bv_2^T,\bv_3^T)^T
+ \boldsymbol{\overline{x^*}}\,$.
The residual of the frame is then
$\br_i=\bx_i-\boldsymbol{\hat{x_i}}\;$
whose 921,600 components (pixels) we 
can normalize by their scales.
This allows us to keep those pixels
of the frame where the absolute 
normalized residuals exceed a threshold,
and turn the other pixels grey.
For wrapping, this procedure yields
a new video which only contains the men. 
This method has thus succeeded in 
accurately separating the movements 
from the background.

\begin{figure}[!htb]
\centering
\includegraphics[width=0.84\textwidth]
		{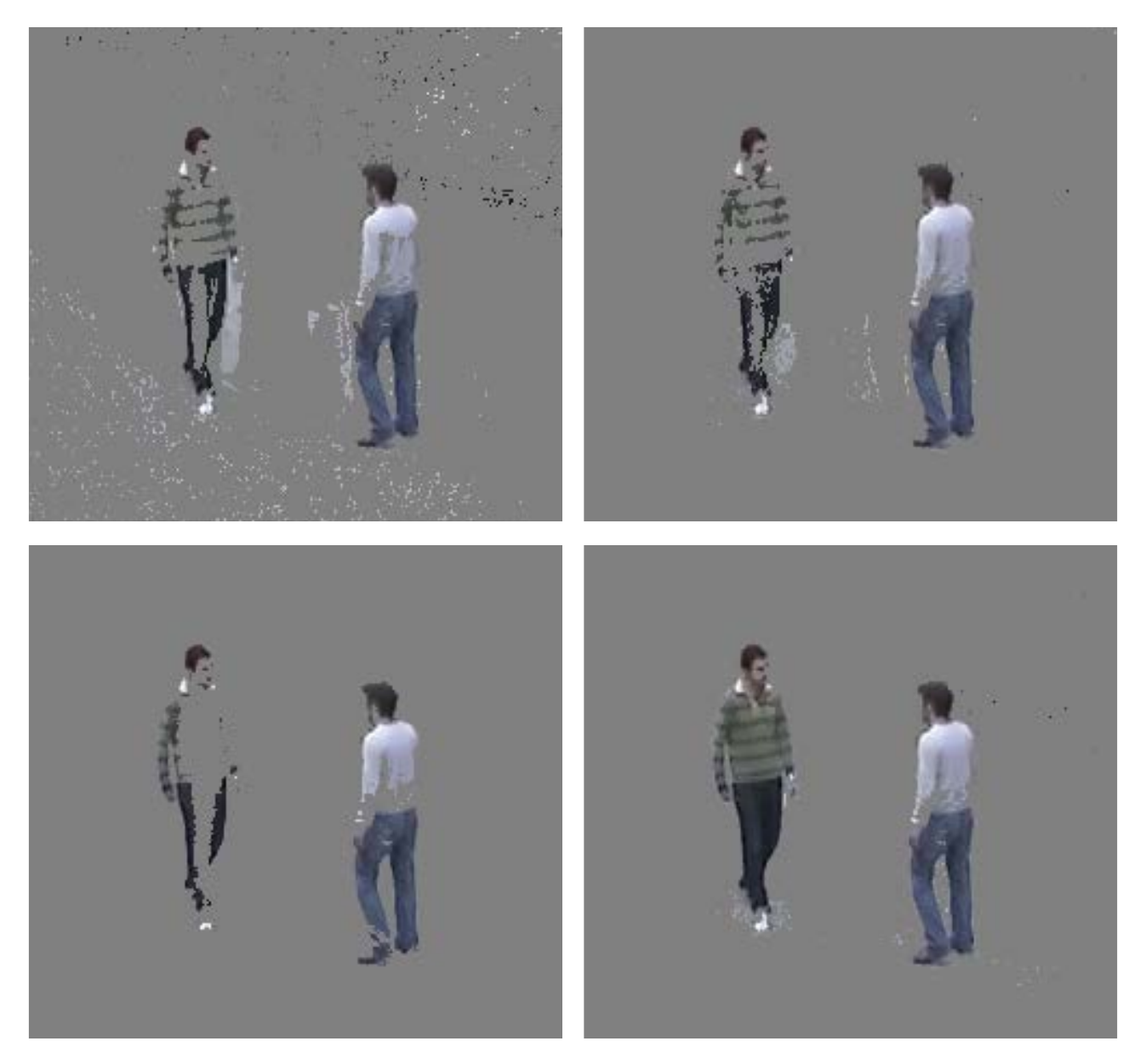} 
\caption{Residuals of the video data, 
for classical PCA (upper left), 
Spearman correlation (upper right),
Huber's $\psi$ (lower left), and
wrapping (lower right).}
\label{fig:mask}
\end{figure}

The lower right panel of Figure
\ref{fig:mask} shows the result for
the central part of frame 100.
The corresponding computation for
classical PCA is shown in the upper
left panel, which has separated the
men less well: many small elements of
the background are marked
as outlying, whereas parts of the man
on the left are missing.
We conclude that in this dataset
wrapping is the most robust, classical 
PCA the least, and the other methods
are in between.

Note that the entire analysis of this 
huge dataset of size 1.6 Gb in R took 
about two minutes on a laptop for
wrapping (the times for the other
three methods were similar).
This is much faster than one would expect
from the computation times in 
Table \ref{tab:times}, which are
quadratic in the dimension since they
calculate the entire covariance matrix.

Of course, in real-time situations one
would estimate the robust loadings on
an initial set of, say, 100 frames and
then process new images while they are
recorded, which is very fast as it only 
requires a matrix multiplication.
In parallel with this the robust loadings 
can be updated from time to time.

\section{Software availability}
The wrapping transform is 
implemented in the R package 
{\it cellWise} \citep{cellWise2019}
on CRAN, which now
also provides the faster version of 
DDC used in the first example.
The package contains two vignettes
with examples. 
The video data of the second example,
its analysis and the video with 
results can be downloaded from\linebreak
{\it https://wis.kuleuven.be/stat/robust/software}\,.

\section{Conclusions}
\label{sec:concl}
Multivariate data often 
contain outlying (anomalous) values, so 
one needs robust methods that can detect
and accommodate such outliers.
The underlying assumption is that the
variables are roughly Gaussian for the
most part, with some possible outliers 
that do not follow any model and could be
anywhere. (If necessary some variables 
can be transformed first, e.g. by taking
their logarithms.)

For multivariate data in low dimensions,
say up to 20, there exist robust scatter
matrix estimators such as the minimum
covariance determinant (MCD) method
that can withstand many rowwise outliers, 
even those that are not visible in the 
marginal distributions.
We recommend to use such high-breakdown
methods when the dimension allows it.
But in higher dimensions these methods
would require infeasible computation
time to achieve the same degree of
robustness, and then we need to resort
to other methods.

It is not easy to construct robust methods 
that simultaneously satisfy the 
independence property, yield positive
semidefinite matrices, and scale
well with the dimension.
We achieve this by transforming the 
data first, after which the usual
methods based on product moments
are applied.

Based on statistical properties such
as the influence function, the 
breakdown value and efficiency we
selected a particular transform called 
wrapping. It leaves over 86\% of the
data intact under normality, which
preserves partial information about 
the data distribution, granularity, 
and the shape of the relation between 
variables.
Wrapping performs remarkably well in 
simulation. It is especially robust 
against cellwise outliers, where it
outperforms typical rowwise robust
methods. This made it possible to
construct a faster version of the
DetectDeviatingCells method.
The examples show that the wrapping 
approach can deal with very high 
dimensional data.\\

\noindent {\bf Supplementary materials.}
These consist of a text with the 
proofs referenced in the paper, and 
an R script that illustrates the
approach and reproduces the examples.\\

\noindent {\bf Funding.}
This research	has been supported by 
projects of Internal Funds KU Leuven.



\clearpage
\pagenumbering{arabic}
%
\appendix
\numberwithin{equation}{section} 
\section{Supplementary Material} \label{sec:A}
\renewcommand{\theequation}
   {\thesection.\arabic{equation}}

Here the proofs of the results are collected.

\subsection{Proof of Proposition \ref{prop:IF}}
\label{A:proofIFT}
We can generate $(X,Y) \sim F_p$ for 
$\rho \gs 0$ by 
\begin{equation}\label{eq:AFrho}
\begin{bmatrix}
X \\
Y
\end{bmatrix} 
= A
\begin{bmatrix}
U \\
V \\
W
\end{bmatrix}
\end{equation}
where $U,V,W$ follow a symmetric unimodal
distribution $G$ and are i.i.d., and
\begin{equation*}
A = 
\begin{bmatrix}
\sqrt{1-\rho} & 0 & \sqrt{\rho} \\
0 & \sqrt{1-\rho} & \sqrt{\rho}
\end{bmatrix}.
\end{equation*}

\noindent For $G=N(0,1)$ the distribution 
of \eqref{eq:AFrho} equals \eqref{eq:Frho}. 
We now obtain 
$\xi(\rho) = E[\psi(u\sqrt{1-\rho}+
 w\sqrt{\rho})\psi(v\sqrt{1-\rho}+
 w\sqrt{\rho})]$.
Since we are interested in $\rho \approx 0$, 
we can use the Taylor expansion (derived 
with $\delta = \sqrt{\rho}$) to obtain
$\psi(u\sqrt{1-\rho}+w\sqrt{\rho}) = 
 \psi(u) + w\sqrt{\rho}\psi'(u) + 
 \frac{w^2\rho}{2}\psi''(u)+o(\rho)$ 
and similarly for the second factor, 
yielding 9 terms of which only one term 
remains, the others being $o(\rho)$ or 
zero since $\psi$ is odd: 
\begin{align*}
\xi(\rho) = E & \left[\psi(u)\left\{ 
 \psi(v)+w\sqrt{\rho} \psi'(v)+
 \frac{w^2\rho}{2}\psi''(v)\right\}
 \right.\\
 &+ w\sqrt{\rho}\psi'(u)\left\{\psi(v)+
  w\sqrt{\rho} \psi'(v) +
	\frac{w^2\rho}{2}\psi''(v)\right\} \\
 &+\left. \frac{w^2\rho}{2} \psi''(u)
  \left\{\psi(v) + w\sqrt{\rho}\psi'(v)+
	\frac{w^2\rho}{2} \psi''(v)
	\right\}\right]\\
 =&\rho E\left[w^2\psi'(u)\psi'(v)\right]+
  o(\rho)\\
 =& \rho E[\psi'(u)]E[\psi'(v)] + o(\rho)
\end{align*}
Therefore $\xi'(0) = E[\psi'(u)]^2$ and we 
obtain $\mbox{IF}((x,y),T,F_0) =
 \psi(x)\psi(y)/E[\psi']^2$.

\subsection{Influence function for 
            general $\rho$}
\label{A:IFgen}

We first consider the non Fisher-consistent
functional $T_{\psi} = E[\psi(X)\psi(Y)]$.
The raw influence function of $T_{\psi}$ under 
the distribution $F_{\rho}$ generated 
as in \eqref{eq:AFrho} is then
$$\mbox{IF}_{raw}((x,y),T_{\psi},F_{\rho}) = 
\psi(x)\psi(y) - E_{F_{\rho}}[\psi(X)\psi(Y)]\;.$$
\begin{proof}
Let $F_{\epsilon} = (1-\epsilon)F_{\rho} + 
\epsilon \Delta_{(x,y)}$. 
Then $$T_{\psi}(F_{\epsilon}) =
 (1-\epsilon) E_{F_{\rho}}[\psi(X)\psi(Y)] + 
 \epsilon E_{\Delta_{(x,y)}}[\psi(X)\psi(Y)]\;.$$ 
Differentiating with respect to $\epsilon$ 
at $\epsilon = 0$ yields 
$-E_{F_{\rho}}[\psi(X)\psi(Y)] + \psi(x)\psi(y)$.
\end{proof}

Now denote the finite sample version of 
$T_{\psi}$ by $T_{n} = \frac{1}{n}
\sum_{i=1}^{n}{\psi(x_i)\psi(y_i)}$.
From the law of large numbers we have that 
$T_{n}$ is strongly consistent for its 
functional value:
$T_{n} \xrightarrow{a.s.} T_{\psi}(F_{\rho})$
for $n \to \infty$. By the central limit
theorem, we also have asymptotic normality of
$T_{\psi}$:
\begin{equation*}
  \sqrt{n}(T_n-T_{\psi}(F_{\rho}))
  \rightarrow N(0,V_{raw})
\end{equation*}
where the asymptotic variance $V_{raw}$ is 
given by
\begin{align*}
 V_{raw} =& \;E_\rho[\mbox{IF}_{raw}((X,Y),
	     T_{\psi},F_{\rho})^2]\\
	=& \;E_\rho\left[\left(\psi(X)\psi(Y) - 
	   E_\rho[\psi(X)\psi(Y)]\right)^2\right]\\
	=& \;E_\rho\left[\psi(X)^2\psi(Y)^2\right] -
	   E_\rho[\psi(X)\psi(Y)]^2 \;\;.
\end{align*}

Now we switch to the Fisher-consistent 
functional $U_{\psi}(F) \coloneqq 
\xi^{-1}(T_{\psi}(F))$ given in
\eqref{eq:xi}.
The general influence function defined in 
\eqref{eq:generalIF} then becomes
\begin{align*}
\mbox{IF}((x,y),T_{\psi},F_{\rho}):=&\; 
\mbox{IF}_{raw}((x,y),U_{\psi},F_{\rho})\\
=&\; \frac{\mbox{IF}_{raw}((x,y),T_{\psi},F)}
	{\xi'(\rho)}\\
=&\; \frac{\psi(x)\psi(y) - 
  E_\rho[\psi(X)\psi(Y)]}{\xi'(\rho)}
\end{align*}
hence
\begin{equation} \label{eq:IFrho}
  \mbox{IF}((x,y),T_{\psi},F_{\rho}) =
  \frac{\psi(x)\psi(y) - C_\rho}{D_\rho}
\end{equation}
where 
$C_\rho \coloneqq E_\rho[\psi(X)\psi(Y)]$ 
and $D_\rho \coloneqq \xi'(\rho)$ can be 
computed numerically
to any given precision. For $\rho=0$ this 
simplifies to the formula in Proposition 1.
Note that the influence function has the
same shape for all values of 
$\rho$ (including $\rho=0$), only the 
constants $C_\rho$ and  $D_\rho$ differ 
which amounts to shifting and rescaling 
the IF along the vertical axis.\\

Now consider the estimator 
$T_{n}^* = \xi^{-1}(T_n)$ corresponding
to the functional $U_\psi$\,.
Since $T_n$ is asymptotically normal, we 
can apply the delta method to establish 
the asymptotic normality of $T_{n}^*$\,. 
Using $(\xi^{-1}(x))'=1/\xi'(\xi^{-1}(x))$ 
we obtain
\begin{equation*}
  \sqrt{n}(T_n^{*}-\rho)
	\rightarrow N\left(0,V \right)
\end{equation*}
where $V = V_{raw}/(\xi'(\rho))^2$ with 
$V_{raw}$ as above.
At $\rho = 0$ this corresponds to
\eqref{eq:V}.

\subsection{Relation with influence functions
            of rank correlations}
\label{A:rankIF}
At the model distribution $F_0$ of 
\eqref{eq:Frho} the influence functions of 
the Quadrant and Spearman correlation
\citep{Croux:IFspearman} and the normal scores
\citep{Boudt:GRcor} correspond to those of
certain $\psi$-product moments.
This is not a coincidence, because 
if we write the rank transform as
$g(x_i) = h(R_n(x_i))$ it tends to the
function $\tilde{g}(x) = h(\Phi(x))$ when 
$n \rightarrow \infty$.
If we put $\psi(x) := h(\Phi(x))$ we observe
that \eqref{eq:splitinf} indeed holds, with
$\mbox{IF}(x,h,\Phi) =
 h(\Phi(x))/\int{(h(\Phi))'d\Phi} =
 \psi(x)/E[\psi']$.

For the quadrant correlation 
$h(u) = \sign(u - 1/2)$ we get the IF of the median: 
\begin{equation*}
\mbox{IF}(x,L_h,\Phi) = 
    \frac{\sign(x)}{2\Phi'(0)}=
    \sqrt{\frac{\pi}{2}}\sign(x)
\end{equation*}
and so $\gamma^{*} = \pi/2$ 
and $\mbox{eff} = 4/\pi^2$.

For the normal scores rank correlation we 
have $h(u) = \Phi^{-1}(u)$ hence
$\mbox{IF}(x,L_h,\Phi) = x$ which is the influence 
function of the mean and thus unbounded, yielding
$\gamma^{*} = \infty$ and $\mbox{eff} = 1$.
The truncated normal scores
$h(u) = \Phi^{-1}\left([u]_{\alpha}^{1-\alpha}\right)
 = [\Phi^{-1}(u)]_{-b}^{b}$ where 
$\alpha = \Phi(-b)$ yields 
$\mbox{IF}(x,L_h,\Phi) =
  \psi_b(x)/E[\psi_b']$, 
which is the influence function of Huber's 
$\psi_b$ function. 

For the Spearman correlation 
($h(u) = u-1/2$) we obtain
\begin{equation*}
  \mbox{IF}(x,L_h,\Phi) = 
	\frac{\Phi(x)- 1/2}{E[(\Phi')^2]} = 
	2\sqrt{\pi} \left(\Phi(x)-\frac{1}{2}\right)
\end{equation*}
which is also the influence function of the 
Hodges-Lehmann estimator and the Mann-Whitney 
and Wilcoxon tests \citep{Hampel:IFapproach}.
It yields $\gamma^{*} = \pi$ 
and $\mbox{eff} = 9/\pi^2$.

\subsection{Proof of
   Proposition \ref{prop:corbias}
   and Corollary \ref{prop:breakdown}}
\label{A:proofbias}

{\it Proof of Proposition \ref{prop:corbias}.}
We give the proof for the maximum upward bias
(the result for the maximum downward bias then
follows by replacing $Y$ by $-Y$).
The uncontaminated distribution of $(X,Y)$ is 
$F =F_\rho$ from \eqref{eq:AFrho}. 
Since $\psi(X)$ and $\psi(Y)$ have the same
distribution and $\psi$ is odd and bounded
we find
$E_{F}[\psi(X)] = E_{F}[\psi(Y)] = 0$ 
and
$E_{F}[\psi(X)^2] = E_{F}[\psi(Y)^2]$\,.
Now consider the contaminated distribution 
$G = (1-\eps)F_\rho + \eps H$ where 
$H$ is any distribution.
At $G$ we obtain
\begin{equation*} 
  \Cor_G(\psi(X),\psi(Y)) =
\frac{E_G[(\psi(X)-E_G[\psi(X)])
     (\psi(Y)-E_G[\psi(Y)])]}
{\sqrt{E_G[(\psi(X)-E_G[\psi(X)]^2)]
     E_G[(\psi(Y)-E_G[\psi(Y)])^2]}}
\end{equation*}		
which works out to be
\begin{equation}\label{eq:CorrG}
\frac{(1-\eps)\Cov_F(U,V) + \eps E_H[UV]
    -\eps^2  E_H[U] E_H[V]}
	{\sqrt{((1-\eps)\VF + \eps E_H[U^2]
	  -\eps^2 E_H[U]^2) 
		((1-\eps)\VF+\eps E_H[V^2]
		-\eps^2 E_H[V]^2) }}
\end{equation}
where we denote $U := \psi(X)$ and
$V := \psi(Y)$ to save space, as well as
$\VF := \Var_F(U) = E_F[\psi(X)^2] =
 E_F[\psi(Y)^2] = \Var_F(V)$.

We will show the proof for $\rho = 0$ which
implies that $U$ and $V$ are independent
hence $\Cov_F(U,V)=0$ as this reduces the 
notation, but the proof remains valid if the 
term $(1 - \eps)\Cov_F(U,V) = 
(1 - \eps)\VF T_\psi(F)$ is kept. 
The proof consists of two parts. We first show 
that the contaminated correlation \eqref{eq:CorrG}
is bounded from above by 
\begin{equation}\label{eq:Ceps}
  C(\eps) := \frac{\eps M^2}
   {(1-\eps)\VF+\eps M^2}
\end{equation}
and then we provide a sequence of contaminating 
distributions $H_n$ for which \eqref{eq:CorrG}
tends to this upper bound.

1. Suppose first that 
 $E_H[U]E_H[V] \ls 0$. 
Then we have for the numerator 
of \eqref{eq:CorrG}:
\begin{align*}
  E_H[UV]-\eps E_H[U] E_H[V] &\ls
	E_H[UV]- E_H[U] E_H[V] \\
  &\ls \sqrt{( E_H[U^2]-E_H[U]^2)
	( E_H[V^2]-E_H[V]^2)}\;\;.
\end{align*}
Now consider the denominator of 
\eqref{eq:CorrG} and note that 
\begin{align*}
  \sqrt{((1-\eps)\VF+
	\eps( E_H[U^2]-\eps E_H[U]^2) )
	((1-\eps)\VF+
	\eps( E_H[V^2]-\eps E_H[V]^2)) } \gs\\
  \sqrt{((1-\eps)\VF+
	\eps( E_H[U^2]-E_H[U]^2) ) 
	((1-\eps)\VF+\eps( E_H[V^2]-E_H[V]^2)) }
\end{align*}
because $E_H[U^2]-E_H[U]^2 \gs 0$, 
$ E_H[U^2] \gs 0$, $E_H[U]^2\gs 0$ 
and $0 \ls \eps \ls 1$. 
Therefore, we can bound \eqref{eq:CorrG} from 
above by 
\begin{equation*}
 \frac{\eps \sqrt{( E_H[U^2]-E_H[U]^2)
 ( E_H[V^2]-E_H[V]^2)}}
 {\sqrt{((1-\eps)\VF+
  \eps( E_H[U^2]-E_H[U]^2) ) 
	((1-\eps)\VF+
	\eps( E_H[V^2]- E_H[V]^2)) }}
\end{equation*}
and this quantity is maximal when 
$(E_H[U^2]-E_H[U]^2)$ and 
$( E_H[V^2]-E_H[V]^2)$ are as large as 
possible. Their supremum is in fact $M^2$.
Therefore, \eqref{eq:CorrG} is less than
or equal to \eqref{eq:Ceps}. 

2. Suppose now that $E_H[U]E_H[V]>0$. 
We will first show that the numerator is 
bounded as follows:
\begin{equation}\label{eq:toprove}
  E_H[UV]-\eps E_H[U] E_H[V] \ls 
  \sqrt{(E_H[U^2]-\eps E_H[U]^2)
	(E_H[V^2]-\eps E_H[V]^2)} \;\;.
\end{equation}
By squaring both sides we find that this
is equivalent to showing
\begin{align*}
E_H[UV]^2-2\eps E_H[U]E_H[V]E_H[UV]\\
\ls E_H[U^2]E_H[V^2]-
\eps(E_H[U^2]E_H[V]^2+E_H[U]^2E_H[V^2])
\end{align*}
which is equivalent to 
\begin{equation}\label{eq:big}
 E_H[U^2]E_H[V^2]-E_H[UV]^2 +
 \eps (2E_H[U]E_H[V]E_H[UV]-
 E_H[U^2]E_H[V]^2-E_H[U]^2E_H[V^2]) \gs 0.
\end{equation}
We know that \eqref{eq:toprove} holds 
for $\eps = 1$ as it is equivalent to
$\Cov_H(U,V) \ls \sqrt{\Var_H(U)\Var_H(V)}$
so \eqref{eq:big} is true in that case.

The general version of \eqref{eq:big} with
$\eps \ls 1$ equals the LHS for $\eps = 1$,
plus $(1-\eps)$ times
\begin{equation}\label{eq:positive}
 E_H[U]^2E_H[V^2]-2E_H[U]E_H[V]E_H[UV]+
   E_H[U^2]E_H[V]^2 \;\;.
\end{equation}
Therefore, it would suffice to prove that
\eqref{eq:positive} is nonnegative.
We know that 
$|E_H[UV]| \ls \sqrt{E_H[U^2]E_H[V^2]}$ 
by Cauchy-Schwarz.
Since $E_H[U]E_H[V]>0$ we obtain
\begin{align*}
E_H[U]^2E_H[V^2]-2E_H[U]E_H[V]E_H[UV]+
  E_H[U^2]E_H[V]^2 \\
 \gs E_H[U]^2E_H[V^2]-2E_H[U]E_H[V]
  \sqrt{E_H[U^2]E_H[V^2]}+
	E_H[U^2]E_H[V]^2 \\
 = \left(E_H[U]\sqrt{E_H[V^2]}-E_H[V]
   \sqrt{E_H[U^2]}\right)^2 \gs 0\;\;.
\end{align*}

Now that we have shown \eqref{eq:toprove}
we can proceed as in part 1,
since \eqref{eq:CorrG} is bounded from 
above by
\begin{equation*}
 \frac{\eps 
 \sqrt{( E_H[U^2]- \eps E_H[U]^2)
 ( E_H[V^2]- \eps E_H[V]^2)}}
 {\sqrt{((1-\eps)\VF+
  \eps( E_H[U^2]- \eps E_H[U]^2) ) 
	((1-\eps)\VF+
	\eps( E_H[V^2]- \eps E_H[V]^2)) }}
\end{equation*}
and this quantity is maximal when 
$(E_H[U^2]- \eps E_H[U]^2)$ and 
$( E_H[V^2]- \eps E_H[V]^2)$ are as large 
as possible. Their supremum is again 
$M^2$, so \eqref{eq:CorrG} is less 
than or equal to \eqref{eq:Ceps}. 

3. Now all that is left to show is that 
the upper bound \eqref{eq:Ceps} is sharp.
Let $(k_n)_{n\in \mathbb{N}}$ be a sequence 
such that 
$\lim_{n \to \infty} \psi(k_n) =
 \sup_x |\psi(x)| = M$ and consider
the sequence of `worst-placed' 
contaminating distributions 
\begin{equation}\label{eq:Hn}
  H_n = \frac{1}{2} \Delta_{(k_n,k_n)} +
     \frac{1}{2} \Delta_{(-k_n,-k_n)}\;\;.
\end{equation}
For the numerator of \eqref{eq:CorrG}
we have 
$\displaystyle \lim_{n \to \infty}
{\eps E_{H_n}[UV]-\eps^2  
E_{H_n}[U] E_{H_n}[V]} = \eps M^2$
since $E_{H_n}[U] = 0 = E_{H_n}[V]$, 
and for the denominator we obtain 
analogously
\begin{align*}
\displaystyle \lim_{n \to \infty}
  \sqrt{((1-\eps)\VF+\eps E_{H_n}[U^2])
      ((1-\eps)\VF+\eps E_{H_n}[V^2])}
  \;=\; (1-\eps)\VF+\eps M^2
\end{align*}
so we reach the upper bound
\eqref{eq:Ceps}.
The proof for the maximum downward bias is 
entirely similar, and there the 
worst placed contaminating distributions
are of the form
$H_n = \frac{1}{2} \Delta_{(k_n,-k_n)} +
 \frac{1}{2} \Delta_{(-k_n,k_n)}$\;. 
QED.\\

{\it Proof of Corollary 
     \ref{prop:breakdown}.}
For the breakdown value we start from 
$F=F_1$\;, that is $\rho=1$ and $X=Y$, so
$\Cov_F(\psi(X),\psi(Y)) = \Var_F(\psi(X))$ 
hence $T_\psi(F) = 1$.
From Proposition \ref{prop:corbias} we know
that 
$$\inf_{G \in \mathcal{F}_\eps}
	 T_\psi(G)
   = \frac{(1-\eps)
	\Var_F(\psi(X))\,T_\psi(F) - \eps M^2}
	{(1-\eps)\Var_F(\psi(X)) +
	\eps M^2}\;\;.$$
For this to be nonpositive the numerator
has to be, i.e.
$(1-\eps)\Var_F(\psi(X)) - \eps M^2 \ls 0$.
The smallest $\eps$ for which this holds
is indeed 
$\Var_F(\psi(X))/(\Var_F(\psi(X))+M^2)$\;.
QED.\\

Note that we can rewrite the breakdown 
value as
$\eps^{*} = 1-(E_F[(\psi/M)^2]+1)^{-1}$ 
so it is a strictly increasing function 
of $E_F[(\psi/M)^2]$.
This implies that the maximizer of the 
breakdown value is $\psi(x) = \sign(x)$ 
which maximizes $E_F[(\psi/M)^2] = 1$, 
hence $\eps^{*}=0.5$ (this yields the 
quadrant correlation).
Interestingly, the breakdown value of 
the scale M-estimator $S$ defined by
$\ave_i \rho(x_i/S) = E_F[\rho]$ where 
$\rho(z) := \psi^2(z)$ is also determined 
by the 
ratio $E_F[\rho]/M^2=E_F[(\psi/M)^2]$,
see e.g. \cite{Maronna:RobStat}.

\subsection{Relation with breakdown values
            of rank correlations}
\label{A:rankBD}

The breakdown values of the rank correlations
in Table \ref{tab:corrs} were derived by
\cite{Caperaa:tauxderes} and 
\cite{Boudt:GRcor}, but not for the 
$\eps$-contamination model 
\eqref{eq:epscont}. 
Instead they used {\it replacement
contamination}, which means you can take out
a certain fraction of the observations
and replace them by arbitrary points.
In fact $\eps$-contamination is a special
case of this, which corresponds to replacing
a mass $\eps$ distributed exactly like the 
original distribution $F$, whereas in
general one could replace an arbitrary
part of $F$.
Therefore the breakdown value for 
replacement is always less than or equal
to that for $\eps$-contamination.
However, in many situations the result turns 
out to be the same, as is the case here.

For rank correlations in the replacement 
model, \cite{Caperaa:tauxderes} and 
\cite{Boudt:GRcor} showed that given a
sorted sample 
$(x_1,y_1), \ldots, (x_n,y_n)$ where 
$x_1< \dots < x_n$ and $x_i = y_i$ for 
all $i \in \{1,\ldots,n\}$, the worst 
possible bias is reached by replacing the 
highest and the lowest $y_i$ by values 
beyond the other end of the range.

We can in fact obtain the same type of 
configuration through the 
$\eps$-contamination model.
Let us start from perfectly correlated
data, that is  $x_i = y_i$ for 
all $i \in \{1,\ldots,n\}$.
Then choose a sequence of
contaminating distributions 
$H_n = \frac{1}{2}\Delta_{(-k_n,k_n)}
+\frac{1}{2}\Delta_{(k_n,-k_n)}$ 
in which the $k_n$ are positive and
tend to infinity, so the 
horizontal and vertical coordinates of 
the outliers move outside the range of 
the original data values.
The resulting rank pairs then have the
same configuration as was constructed
for breakdown under replacement.
Therefore the $\eps$-contamination
breakdown values of rank correlations
equal those under replacement.

\subsection{Construction of the optimal
  transformation}
\label{A:wrapping}

Theorem 3.1 in \citep{Hampel:tanh} says that for any
$0<c<\infty$ and large enough $k > 0$ there exist 
positive constants $0<b<c$, $A$ and $B$ such that 
$\tilde{\psi}$ defined by
\begin{equation}\label{eq:oldTanh}
\tilde{\psi}(z) = \begin{cases}
  z &\mbox{ if } 0 \ls |z| \ls b \\
  \sqrt{A(k-1)}\tanh\left(\frac{B}{2}\sqrt
	{\frac{k-1}{A}}(c-|z|)\right)\sign(z)
		&\mbox{ if } b \ls |z| \ls c\\
  0 &\mbox{ if } c \ls |z|
\end{cases}
\end{equation}
satisfies
$$ b = \sqrt{A(k-1)}\tanh\left(\frac{1}{2}
  \sqrt{\frac{(k-1)B^2}{A}}(c-b)\right)\;\;,$$
$A = \int_{-c}^{c}{\tilde{\psi}(x)^2d\Phi(x)}$\;, 
$B =\int_{-c}^{c}{\tilde{\psi}'(x)d\Phi(x)}$
and $\kappa^{*}(\tilde{\psi}) = k$\;. 
Theorem 4.1 then says that this function 
$\tilde{\psi}$ minimizes the asymptotic variance
among all odd functions $\psi$ satisfying
\eqref{eq:redesc} subject to 
$\kappa^{*}(\psi) \ls k$,
and that this optimal solution is unique (upto
a positive nonzero factor).
It can be verified that for a given value of $c$
there is a strictly monotone relation between
$k$ and $b$, so we have decided to parametrize
$\tilde{\psi}$ by the easily interpretable 
tuning constants $b$ and $c$.
A short R-script is available that for any
$b$ and $c$ derives the other constants
$A$, $B$ and $k$, in turn yielding 
$q_1=\sqrt{A(k-1)}$ and
$q_2=(B/2)\sqrt{(k-1)/A}$\;. 
For instance, for $b=1.5$ and $c=4$ we 
obtain $A = 0.7532528$, $B = 0.8430849$ 
and $k = 4.1517212$ hence 
$q_1 = 1.540793$ and 
$q_2 = 0.8622731$, yielding the
gross-error-sensitivity 
$(b/B)^2 = 3.16$ and the
efficiency $(B^2/A)^2 = 0.890$.\\

\begin{figure}[!ht]
\centering
\includegraphics[width=0.75\textwidth]
                {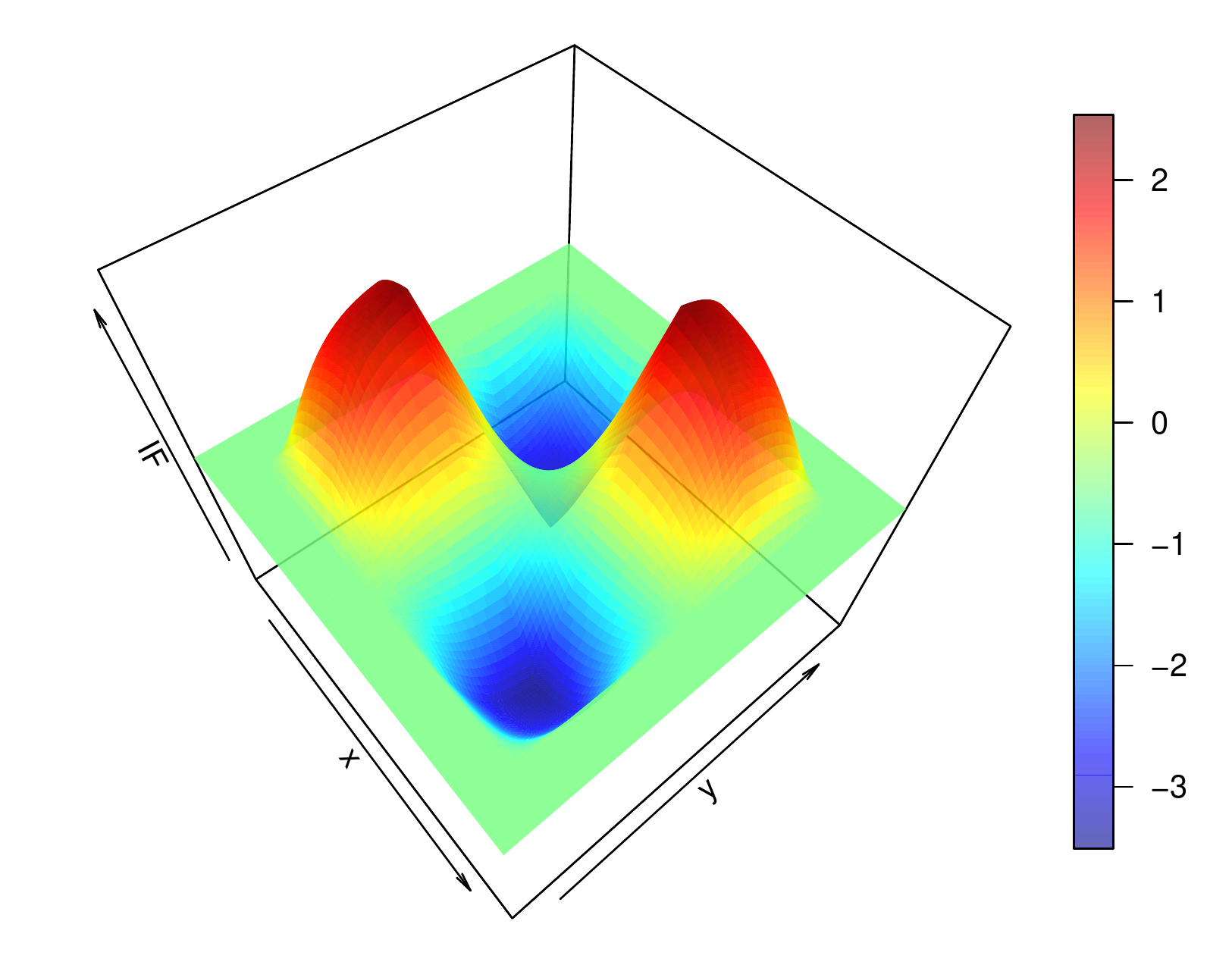}
\vskip-0.2cm					
\caption{Influence function of $T_{\psi}$ 
         at $F_{\rho}$ for $\rho = 0.5$.}
\label{fig:IF05}
\end{figure}

Figure \ref{fig:IF05} shows the influence 
function \eqref{eq:IFrho} at $\rho = 0.5$ 
for the psi-function $\psi_{b,c}$
of \eqref{eq:psiwrap}.
The influence function has the same shape 
at other values of $\rho$, up to shifting
and rescaling the surface along the 
vertical axis, as shown in Section 
\ref{A:IFgen}.\\

\subsection{Proof of
   Propositions \ref{prop:independence}
   and \ref{prop:linearity}}
\label{A:independence}

{\it Proof of Proposition 
\ref{prop:independence}.}
It is assumed that $(X,Y)$ follows a 
bivariate Gaussian distribution.
Due to the invariance properties of 
correlation, we can assume w.l.o.g. that 
the distribution is $F_{\rho}$ with center 0, 
unit variances and true correlation 
$-1 < \rho < 1$.
The assumption that $\Cor(g_X(X),g_Y(Y))=0$
is equivalent to its numerator being
zero, i.e. 
$T(F_{\rho}) = E_\rho[\psi(X)\psi(Y)] = 0$.
We need to show that this implies 
$\rho = 0$, from which independence between 
the components follows.

We first show that $\rho > 0$ implies that
$T(F_{\rho}) = E_\rho[\psi(X)\psi(Y)] > 0$. 
Denote 
$A = \{(x,y)\in \mathbb{R}^2;\; xy > 0\}$ 
and $B = \{(x,y)\in \mathbb{R}^2 ;\; xy < 0\}$.
We then have:
\begin{align*}
E_\rho[\psi(X)\psi(Y)] =& \int_{\mathbb{R}^2}
     {\psi(x)\psi(y)f_{\rho}(x, y) dxdy}\\
=& \int_{A}{\psi(x)\psi(y)f_{\rho}(x, y) dxdy} + 
    \int_{B}{\psi(x)\psi(y)f_{\rho}(x, y) dxdy}\\
=& \int_{A}{\psi(x)\psi(y)f_{\rho}(x, y) dxdy} +
  \int_{A}{\psi(x)\psi(-y)f_{\rho}(x, -y) dxdy}\\
=& \int_{A}{\psi(x)\psi(y)f_{\rho}(x, y) dxdy} -
 \int_{A}{\psi(x)\psi(y)f_{\rho}(x, -y) dxdy}\\
=& \int_{A}{\psi(x)\psi(y)\left\{f_{\rho}(x, y)
  - f_{\rho}(x, -y)\right\}dxdy}\;.
\end{align*}
In the third equality we have changed the 
integration variables from $(x, y)$ to $(x, -y)$.
This transformation has Jacobian 1 and maps
$B$ to $A$. In the fourth equality we have used 
that $\psi$ is odd so $\psi(-y) = -\psi(y)$. 
Now note that $f_{\rho}(x,y) > f_{\rho}(x,-y)$ 
for all $(x,y) \in A$ since $\rho > 0$. 
We conclude that $T(F_{\rho})>0$.
The proof that $T(F_{\rho}) < 0$ for $\rho < 0$
follows by symmetry. Therefore, $T(F_{\rho}) = 0$ 
implies $\rho = 0$\,.\\

{\it Proof of Proposition 
\ref{prop:linearity}.}

(i) From \eqref{eq:linear} and equivariance
it follows that
$\hmu_Y = \alpha + \beta \hmu_X$ and
$\hs_Y = \beta \hs_X$ hence 
$g_Y(y_i) = 
 (y_i - \hmu_Y)/\hs_Y =
 (x_i - \hmu_X)/\hs_X =
 g_X(x_i)$ for all $i$.

(ii) From $\Cor(g_X(x_i),g_Y(y_i))=1$
and $\ave_i(g_X(x_i))=0$ and
$\ave_i(g_Y(y_i))=0$ it follows that there
is a constant $\gamma >0$ such that
$g_Y(y_i) = \gamma g_X(x_i)$ for all $i$.
For the $i$ for which
$|x_i - \hmu_X|/\hs_X \leqslant b$ and
$|y_i - \hmu_Y|/\hs_Y \leqslant b$
it holds that 
$g_Y(y_i) = (y_i - \hmu_Y)/\hs_Y$ and
$g_X(x_i) = (x_i - \hmu_X)/\hs_X$ hence
$(y_i - \hmu_Y)/\hs_Y = \gamma
 (x_i - \hmu_X)/\hs_X$ 
which implies \eqref{eq:linear} with
$\alpha = \hmu_Y-\gamma\hmu_X\hs_Y/\hs_X$
and $\beta = \gamma \hs_Y/\hs_X$.

\subsection{Illustration of anomaly
detection based on robust location and
scatter}
\label{A:robdist}

\begin{figure}[!ht]
\centering
\includegraphics[width=0.65\textwidth]
                {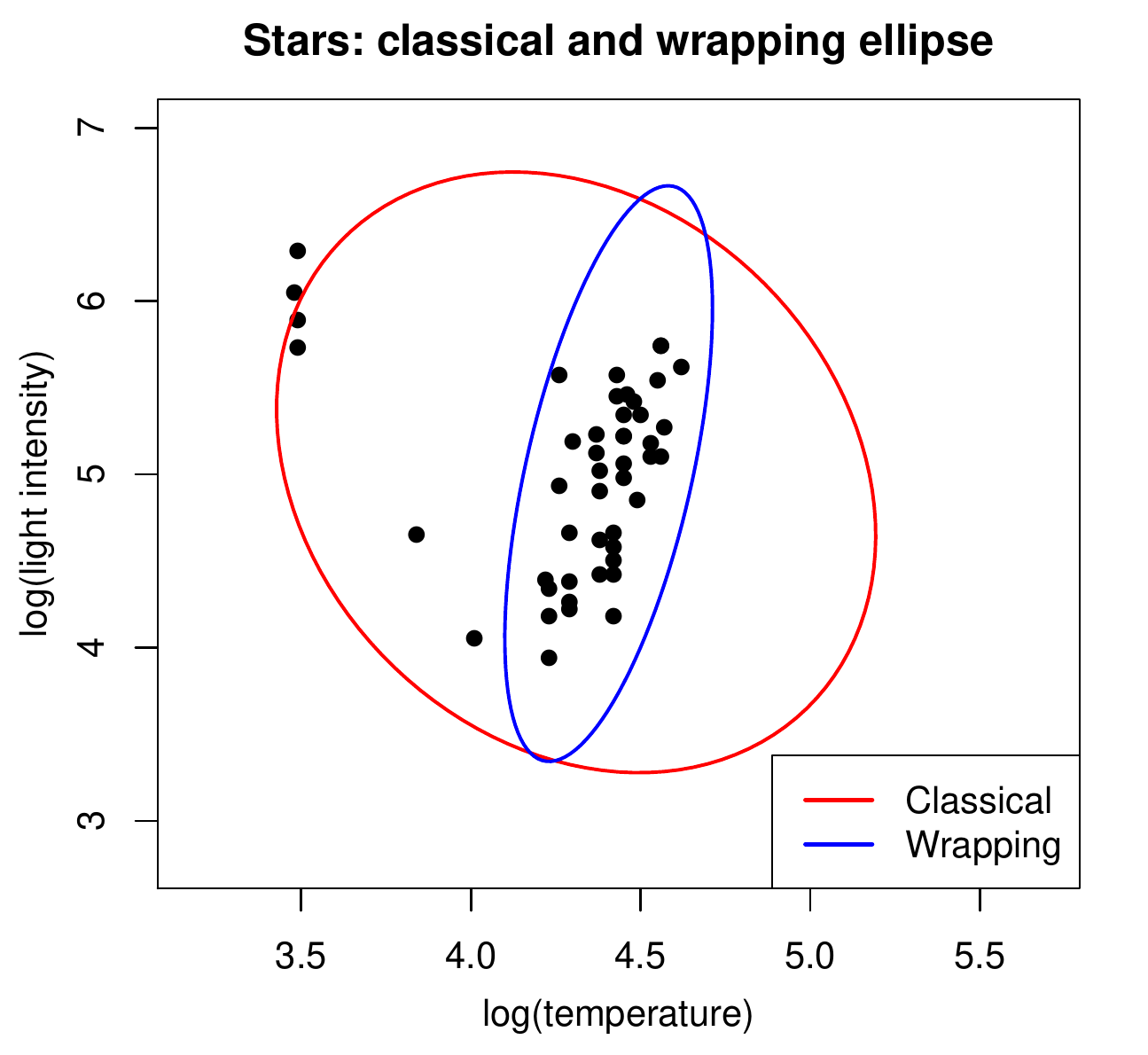}
\vskip-0.2cm						
\caption{Plot of the 47 stars with their
  classical tolerance ellipse (red) and
	the one based on wrapped covariance (blue).}
	\label{fig:stars_ellipses}	
\end{figure}

To visualize things we consider a small
bivariate data set, about the star cluster
CYG OB1 consisting of 47 stars in the direction
of Cygnus. Their Hertzsprung-Russell diagram is
a plot of the logarithm of each star's light 
intensity versus the logarithm of its temperature.
The data can be found on page 27 of \citep{RROD} 
and is plotted in Figure \ref{fig:stars_ellipses}.
We see that the majority of the stars (the 
so-called main sequence stars) follows a certain 
upward trend, whereas there are four anomalous
stars in the upper left corner. These are
red giant stars. 
In this data set the anomalies are measured
correctly, but they belong to a different
population.

The classical correlation between the
variables is $-0.21$ which would indicate a 
negative relation. However, this decreasing 
trend is caused by the four outliers, and without
them the trend would be increasing. Indeed,
the wrapped correlation is $0.57$ indicating
a positive relation. 
Figure \ref{fig:stars_ellipses} shows the $99\%$
tolerance ellipse derived from the classical 
mean and covariance matrix, in red. 
The four outliers have pulled the ellipse toward
them, making them lie on its boundary.
In contrast, the tolerance ellipse from the 
wrapped mean and covariance (in blue) fits the
majority of the stars, leaving aside the four
outliers.

\begin{figure}[!ht]
\centering
\vskip0.5cm
\includegraphics[width=0.49\textwidth]
                {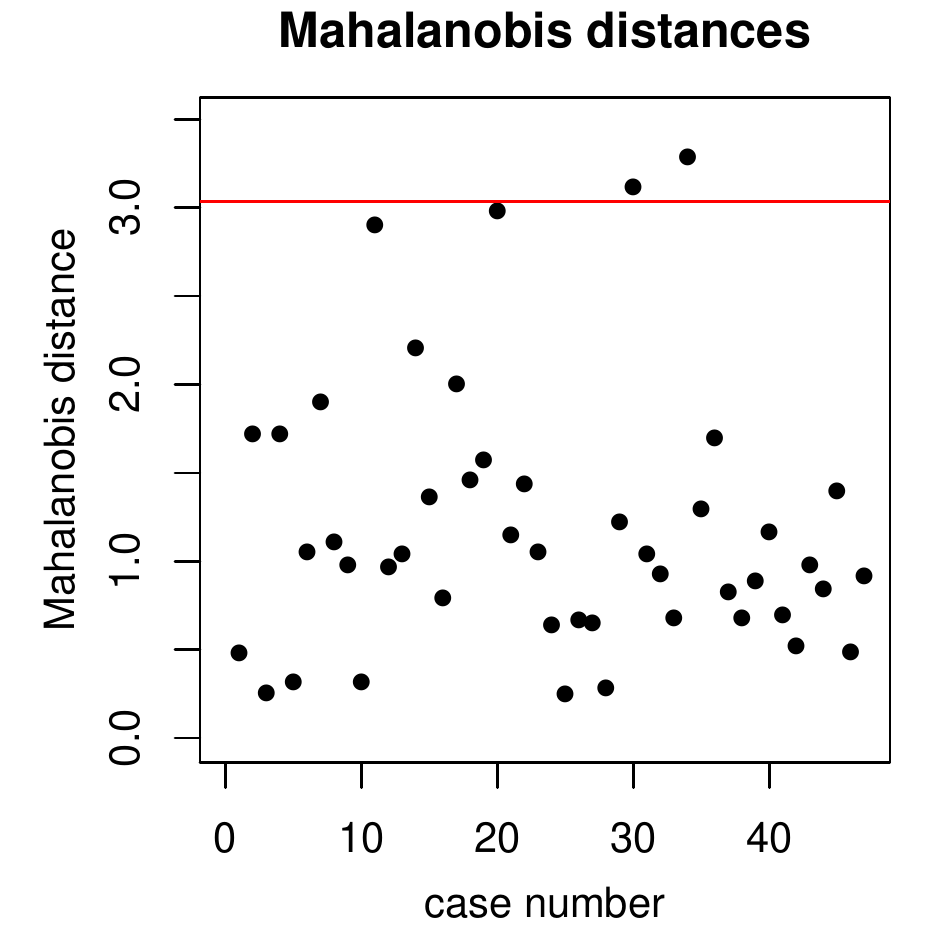}
\includegraphics[width=0.49\textwidth]
                {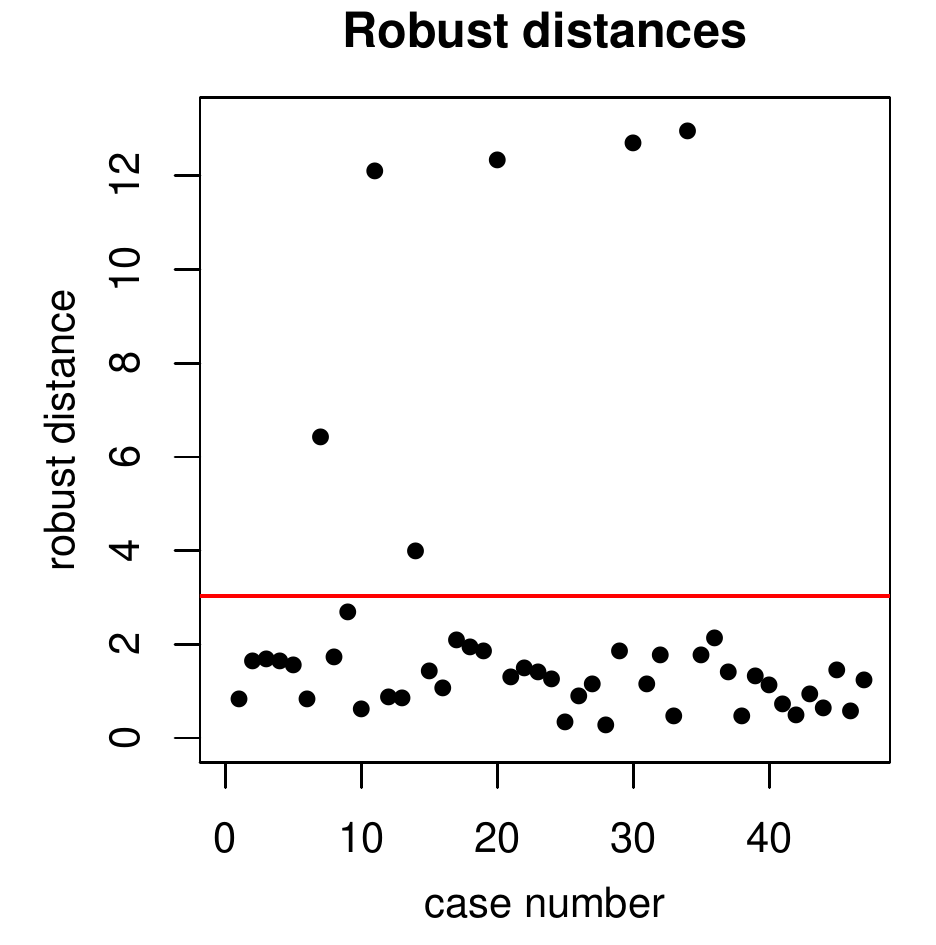}
\vskip-0.2cm						
\caption{Classical distances of the stars
  (left) and their robust distances based
	on wrapped location and covariance (right).}
	\label{fig:stars_distances}	
\end{figure}

Of course, in higher dimensions we can no 
longer plot the data points or draw the
tolerance ellipsoids.
But in that case we can still look at the
classical Mahalanobis distance of each case
$\bx_i$ given by
\begin{equation}\label{eq:MD}
  \mbox{MD}(\bx_i) = \sqrt{(\bx_i - \hbmu)'
	   \hbs^{-1} (\bx_i - \hbmu)}\;\;,
\end{equation}
in which $\hbmu$ is the arithmetic mean and
$\hbs$ the empirical covariance matrix.
The left panel of Figure 
\ref{fig:stars_distances} plots 
$\mbox{MD}(\bx_i)$ versus the case number $i$.
In this plot the four giant stars lie close to
the cutoff value $\sqrt{\chi^2_{d,0.99}}$
for dimension $d=2$.
But they are easily detected in the right hand
panel, which plots the robust distances given
by \eqref{eq:MD} where this time  $\hbmu$ and 
$\hbs$ are the location and scatter matrix 
obtained from the wrapped data. 
These robust estimates have thus allowed us 
to detect the anomalies.

\subsection{Distance correlation
  after transformation}
\label{A:distcov}

The distance correlation $\mbox{dCor}$
between random vectors $\bX$ and $\bY$ is defined 
by the Pearson correlation between the doubly 
centered interpoint distances of $\bX$ and
those of $\bY$ \citep{Szekely:distcor}. 
It always lies between 0 and 1.
Interestingly, $\mbox{dCor}(\bX,\bY)$ can also
be written in terms of the characteristic 
functions of the joint distribution of
$(\bX,\bY)$ and the marginal distributions of
$\bX$ and $\bY$. Using this result 
\cite{Szekely:distcor} prove that
$\mbox{dCor}(\bX,\bY)=0$ implies that $\bX$ 
and $\bY$ are independent, which is not true
for the plain Pearson correlation (except for 
multivariate Gaussian data).

\begin{figure}[!ht]
\centering
\vskip0.5cm
\includegraphics[width=0.49\textwidth]
                {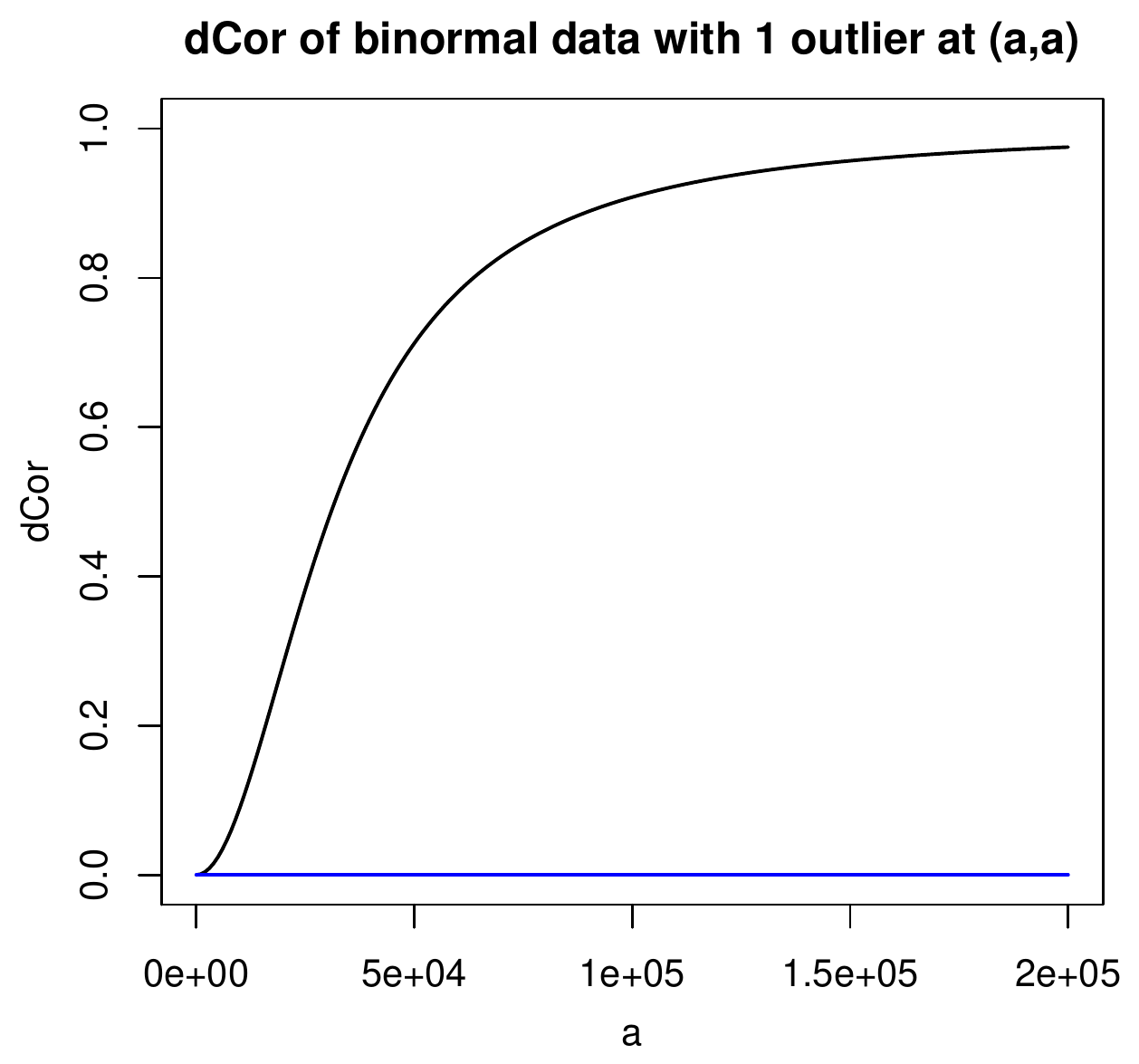}
\includegraphics[width=0.49\textwidth]
                {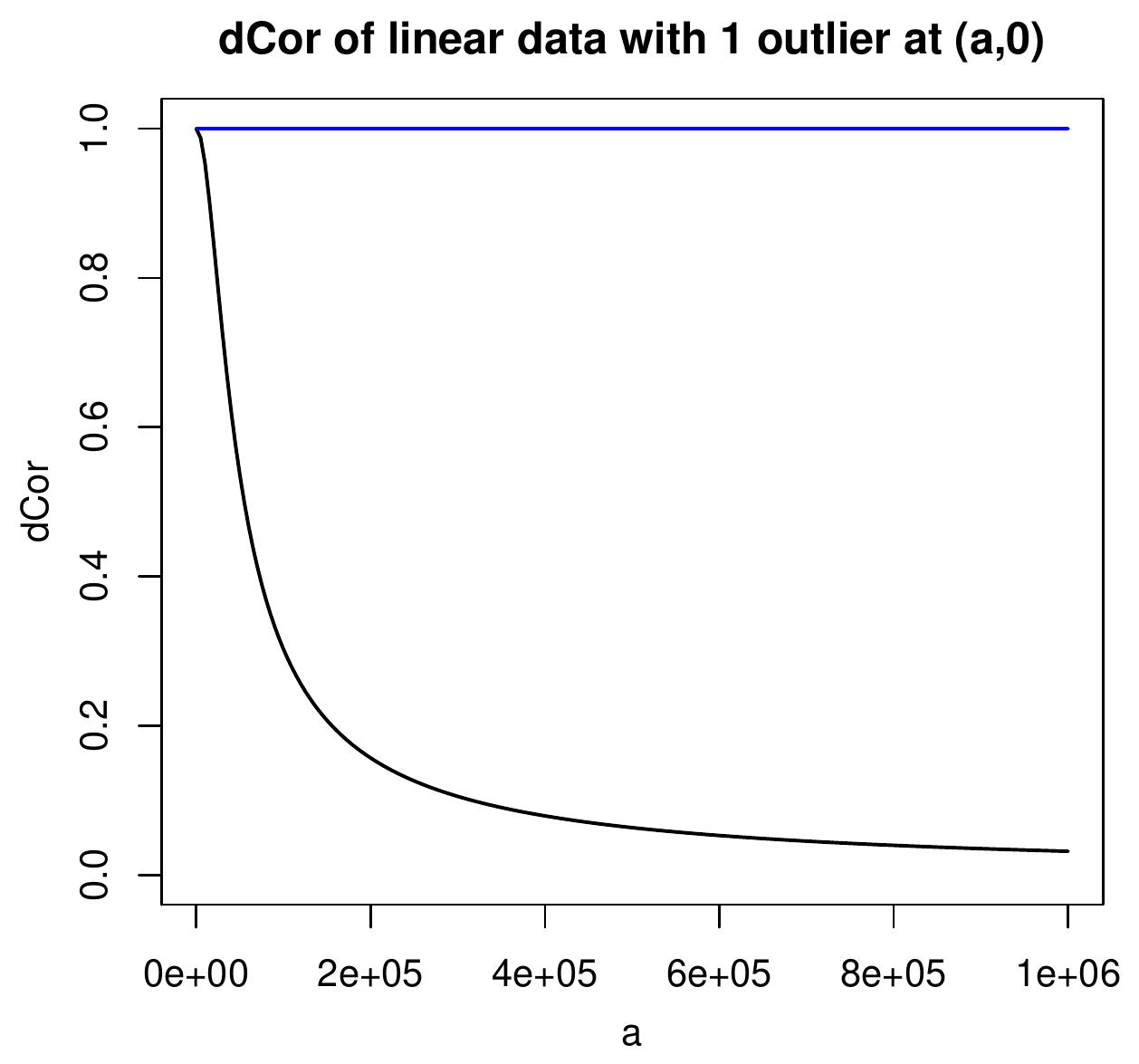}
\vskip-0.2cm						
\caption{Left panel: distance correlation 
(black curve) and its robust version (blue 
curve) of a
data set with $99,999$ standard Gaussian data 
points and one outlier at $(a,a)$ versus $a$.
Right panel: distance correlation of data with
$99,999$ data points $(x_i,x_i)$ with standard
Gaussian $x_i$ and one outlier at $(a,0)$.}
\label{fig:distcor_outlier}	
\end{figure}

The population $\mbox{dCor}(\bX,\bY)$ 
is estimated by its finite-sample version
$\mbox{dCor}(\bX_n,\bY_n)$ which is
a test statistic for dependence.
Unfortunately this statistic is very 
sensitive to outliers.
To illustrate this we first generate
$n=100,000$ data points 
from the standard bivariate Gaussian 
distribution, which has 
$\mbox{dCor}(\bX,\bY)=0$, 
and replace a single observation by an 
outlier in the point $(a,a)$.
The left panel of
Figure \ref{fig:distcor_outlier}
shows $\mbox{dCor}(\bX_n,\bY_n)$ as a 
function of $a$. 
For this we used the fast algorithm of 
\cite{Huo:fast} as implemented in the
function {\it dcor2d} in the R package 
{\it energy}, which can handle such a
large sample size $n$.
For $a=0$ we obtain 
$\mbox{dCor}(\bX_n,\bY_n) \approx 0$ but 
by letting $a$ increase we can bring the 
result close to 1, even though the 
remaining $99,999$ points were generated
independently.

We can also do the opposite, by
starting from a perfectly dependent setting.
For this we generate $\bX_n$ from the 
univariate standard Gaussian distribution,
and take $\bY_n\,:=\,\bX_n$ so that 
$\mbox{dCor}(\bX_n,\bY_n) = 1$.
Then we replace a single observation by an 
outlier in the point $(a,0)$.
In the right panel of
Figure \ref{fig:distcor_outlier}
we now see that we can bring 
$\mbox{dCor}(\bX_n,\bY_n)$ close to 0 by this
single outlier out of $100,000$ data points.

We now apply our methodology of first
transforming the individual variables.
For this we use the function $g$ of 
\eqref{eq:wrapx} where $\hmu_j$ is the 
sample median and $\hs_j$ is the 
median absolute deviation.
For the $\psi$-function we use the
sigmoid $\psi(z) = \tanh(z)$.
After this transformation we compute the 
distance correlation.
This combined method no longer requires the 
first
moments of the original variables to exist
because $\psi$ is bounded, and its population 
version is again zero if and only if the 
original $\bX$ and $\bY$ are independent, 
since $\psi$ is invertible.
The blue lines in Figure
\ref{fig:distcor_outlier} are the result
of applying the combined method, which by
construction is insensitive to the outlier.

The robustness of the proposed method can 
help even when no outliers are added but
distributions are long-tailed, as 
illustrated in 
Figure \ref{fig:distcor_Cauchy}.

\subsection{Simulation with
   cellwise outliers}
\label{A:cell}
This section repeats the simulation in
Section \ref{sec:sim} for cellwise
outliers. 
The clean data are exactly the same,
but now we randomly select data cells
and replace them by outliers following
the distribution $N(k,0.01^2)$ when 
they occur in the $x$-coordinate and 
$N(-k,0.01^2)$ when they occur in the 
$y$-coordinate. 
The simulation was run for 10\%, 20\% 
and 30\% of cellwise outliers, but the 
patterns were similar across 
contamination levels. 

\begin{figure}[!ht]
\centering
\includegraphics[width=0.49\textwidth]
           {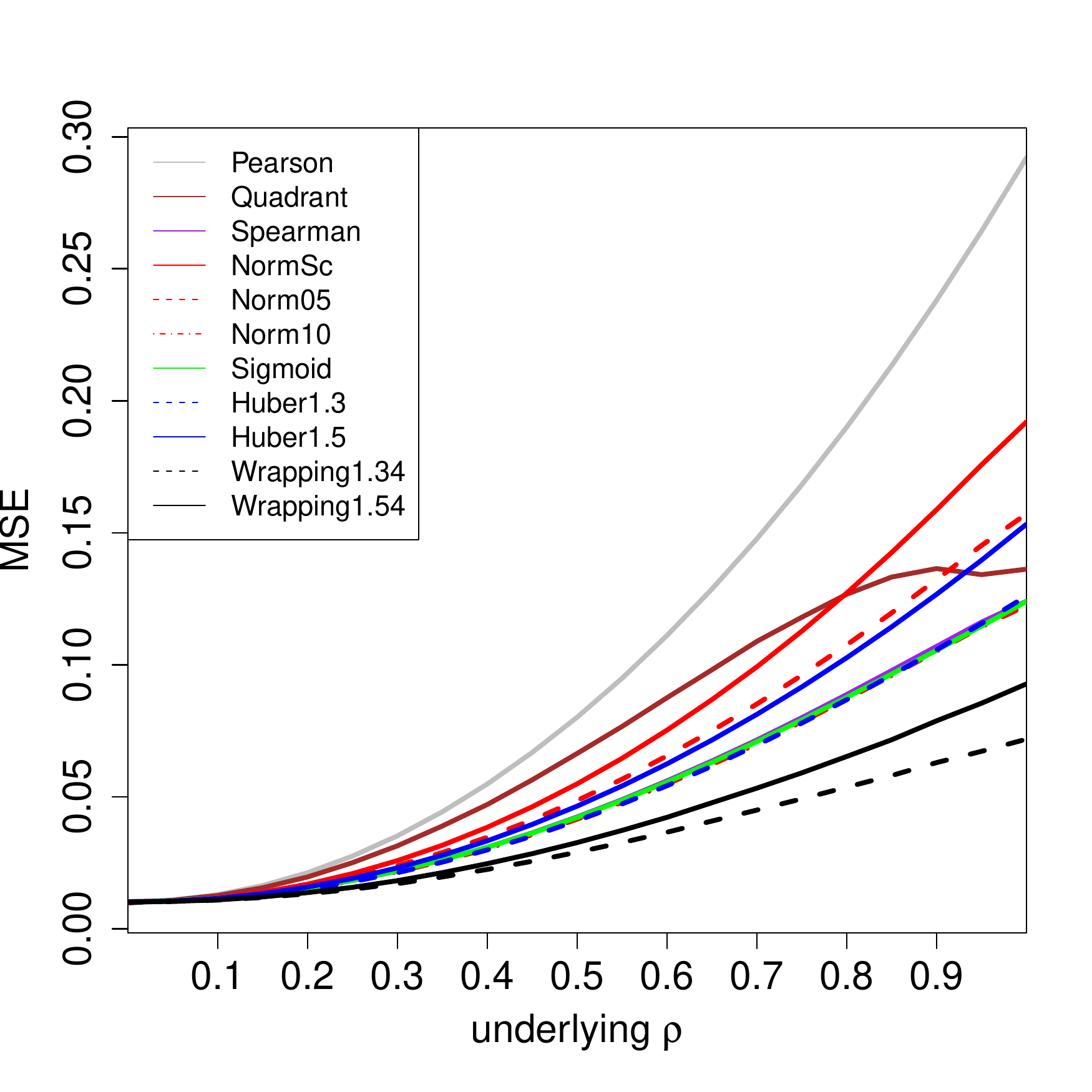}
\includegraphics[width=0.49\textwidth]
           {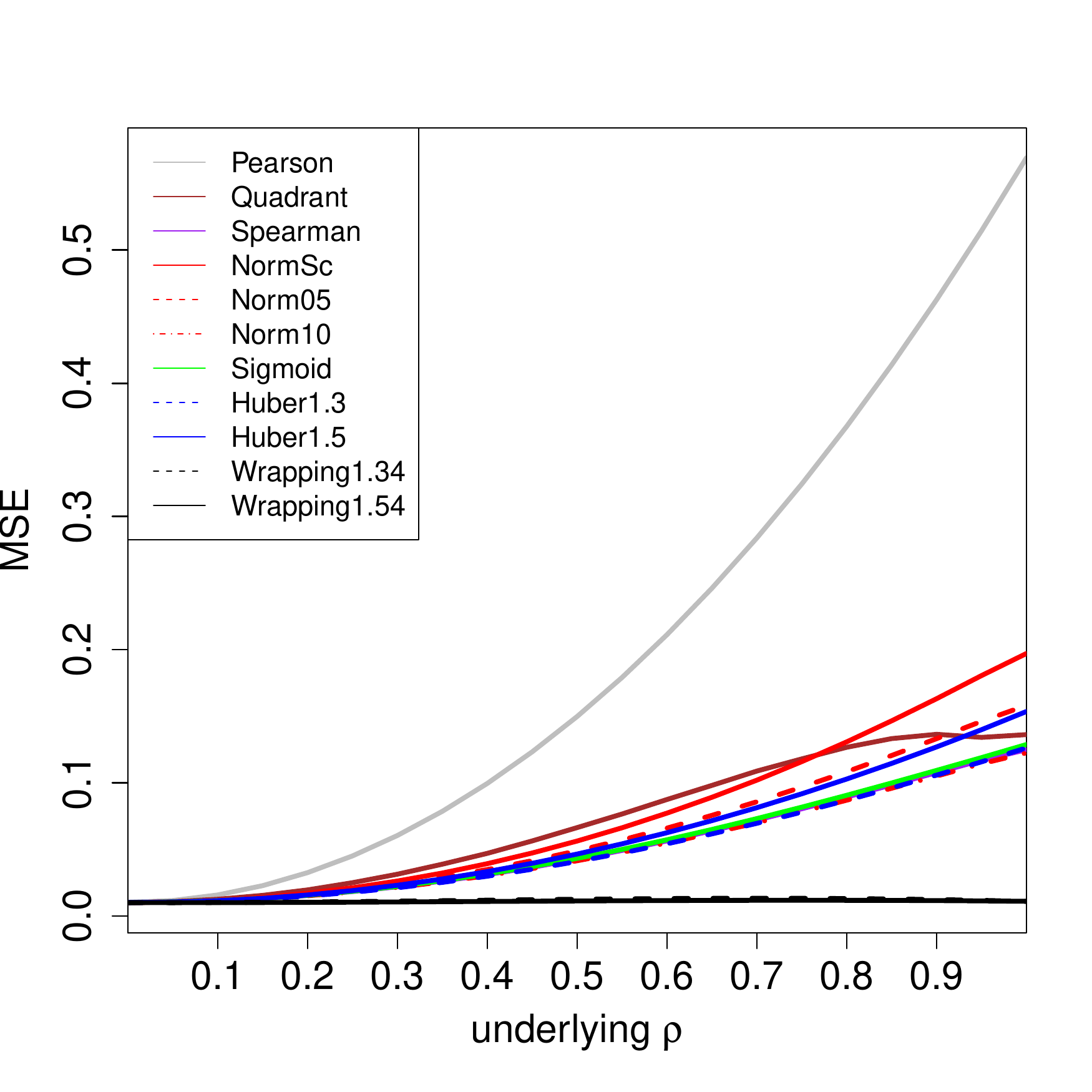}
\vskip-0.2cm						
\caption{MSE of the correlation 
  measures in Figure \ref{fig:CBTclean}
	with 10\% of cellwise outliers placed 
	with $k = 3$ (left) and $k=5$ (right).}
	\label{fig:CBTcell}	
\end{figure}

\begin{figure}[!ht]
\centering
\includegraphics[width=0.49\textwidth]
       {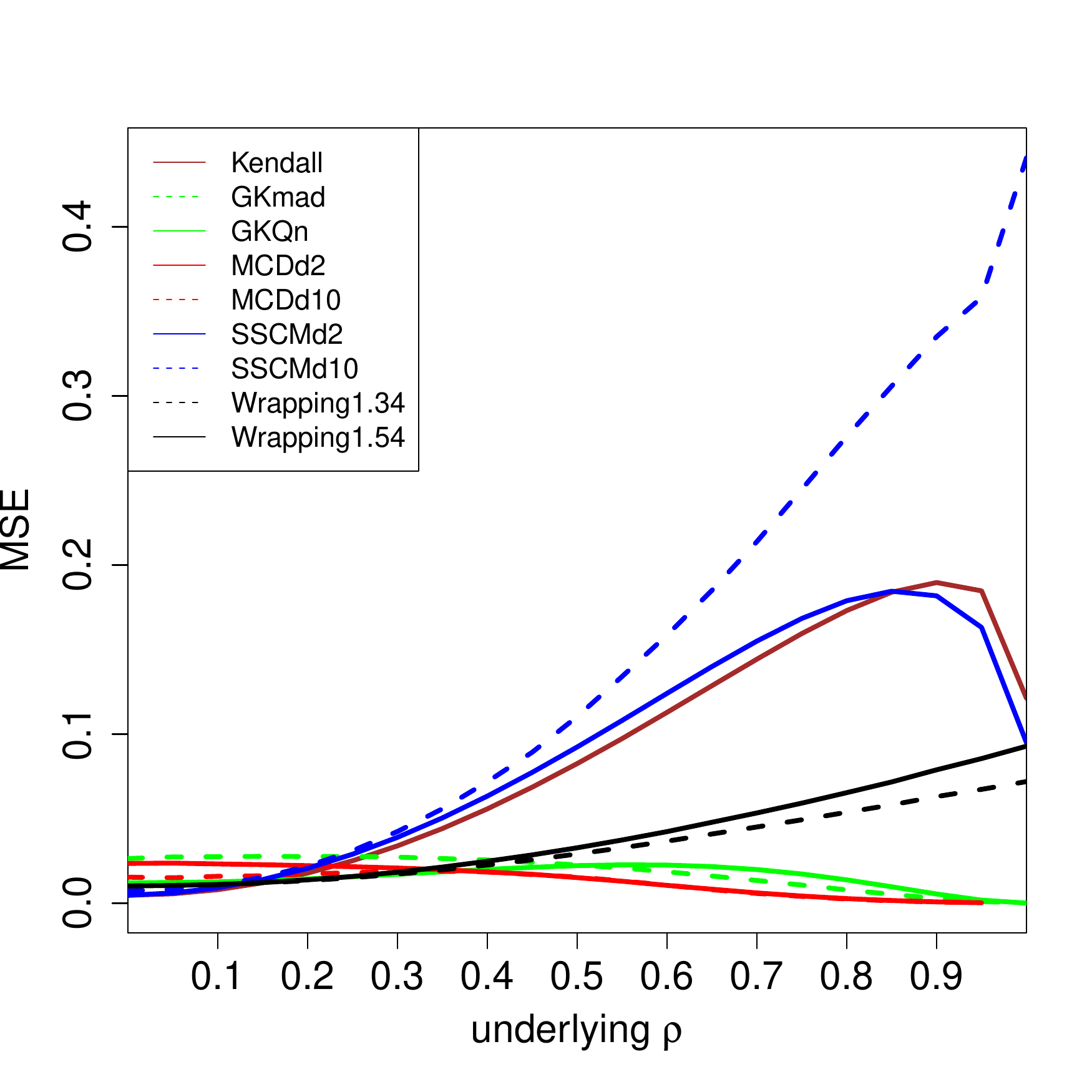}
\includegraphics[width=0.49\textwidth]
       {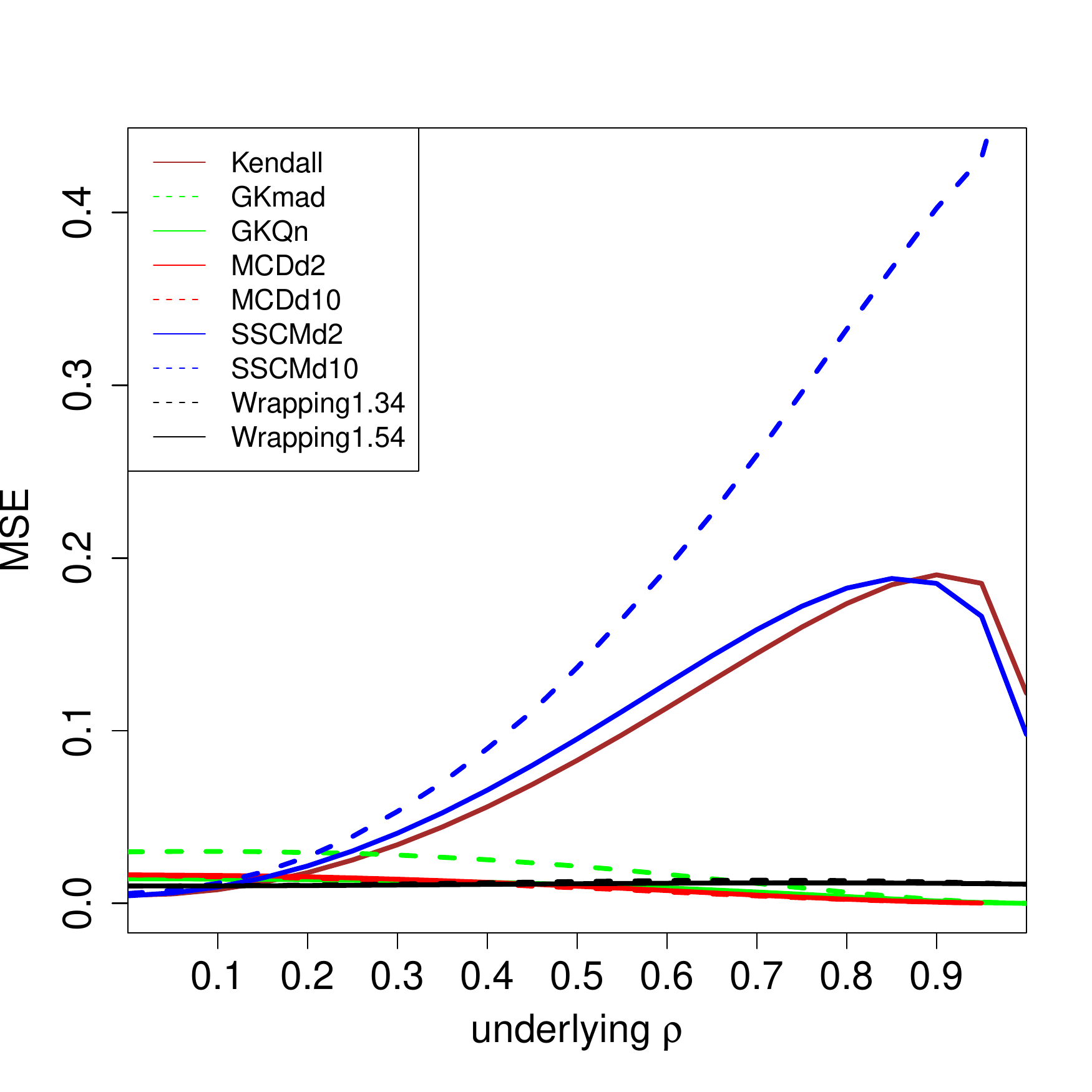}
\vskip-0.2cm						
\caption{MSE of the correlation 
  measures in Figure \ref{fig:CCclean}
	with 10\% of cellwise outliers placed
	with $k = 3$ (left) and $k=5$ (right).}
	\label{fig:CCcell}	
\end{figure}

Figure \ref{fig:CBTcell} shows the MSE
of the same transformation-based
correlation measures as in Figure 
\ref{fig:CBTclean},	with 10\% of 
cellwise outliers for	$k = 3$  
and $k=5$.
Within this class Pearson again has the
worst MSE, followed by normal scores.
The quadrant correlation is next, and
does not look as good here as for 
rowwise outliers.
Wrapping has the lowest MSE, and again
outperforms Spearman, sigmoid and Huber
because it moves the outlying cells to 
the central part of their variable.

Figure \ref{fig:CCcell} compares
wrapping to the correlation measures
in Figure \ref{fig:CCcont} in the
presence of these cellwise outliers.
Also here the SSCM has the largest
bias, especially in $d=10$ dimensions,
followed by Kendall's tau.
Wrapping does well but not as well
as MCD and GK when $k=3$, and 
their performance is similar for $k=5$.
But in higher dimensions wrapping
still has the redeeming feature that
it yields a PSD correlation matrix
unlike the GK method, whereas the MCD
suffers from the propagation
of cellwise outliers and a high
computation time.

\end{document}